\documentclass[preprint2]{aastex}
\renewcommand{\cite}{\citealp}

\usepackage{graphics}
\pdfoutput=1
 
\shorttitle{Hide-and-seek in M31}
\shortauthors{Clementini et al.}

\begin{document}

\title{Hide-and-seek between Andromeda's halo and disk, and giant stream\altaffilmark{1}}
\author{ Gisella Clementini\altaffilmark{2}, Rodrigo Contreras Ramos\altaffilmark{2,3},  
Luciana Federici\altaffilmark{2}, Giulia Macario\altaffilmark{2,3}, Giacomo Beccari\altaffilmark{4}, Vincenzo Testa\altaffilmark{5}, Michele Cignoni\altaffilmark{3}, 
Marcella Marconi\altaffilmark{6}, Vincenzo Ripepi\altaffilmark{6},  
Monica Tosi\altaffilmark{2}, Michele Bellazzini\altaffilmark{2},
Flavio Fusi Pecci\altaffilmark{2}, Emiliano Diolaiti\altaffilmark{2}, Carla Cacciari\altaffilmark{2}, 
Bruno Marano\altaffilmark{3}, 
Emanuele Giallongo\altaffilmark{5}, Roberto Ragazzoni\altaffilmark{7}, Andrea Di Paola\altaffilmark{5}, Stefano Gallozzi\altaffilmark{5}, and 
Riccardo Smareglia\altaffilmark{8}.}
\altaffiltext{1}{Based on data acquired using the blue channel camera of the Large Binocular Telescope (LBT/LBC-blue)} 
\altaffiltext{2}{INAF, Osservatorio Astronomico di Bologna, Bologna, Italy; gisella.clementini@oabo.inaf.it
}
\altaffiltext{3}{Dipartimento di Astronomia, Universit\`a di Bologna, Bologna, Italy}
\altaffiltext{4}{European Southern Observatory, Karl-Schwarzschild-Str. 2, 85748 Garching bei Munchen, Germany}
\altaffiltext{5}{INAF, Osservatorio Astronomico di Roma, Monteporzio, Italy}
\altaffiltext{6}{INAF, Osservatorio Astronomico di Capodimonte, Napoli, Italy}
\altaffiltext{7}{INAF, Osservatorio Astronomico di Padova, Padova, Italy}
\altaffiltext{8}{INAF, Osservatorio Astronomico di Trieste, Trieste, Italy}
\begin{abstract}
Photometry in $B,V$ (down to $V\sim$ 26 mag) is presented for two 
$23^{\prime}\times 23^{\prime}$ fields of the Andromeda galaxy (M31)  
that were observed with the blue channel camera of the Large Binocular 
Telescope during the Science Demonstration Time.
Each field covers  
an area of
about 5.1$\times$5.1 kpc$^2$ at the distance of M31 ($\mu_{\rm M31} \sim 24.4$ mag), 
sampling, respectively, a northeast region close to the M31 giant stream (field S2), and an eastern portion of the 
halo in the direction of
the galaxy minor axis (field H1).
The stream field spans a region that includes Andromeda's disk and the giant
stream, and this is reflected in the complexity of the color magnitude diagram of
the field.  One corner of the halo field also includes a portion of the giant
stream. Even though these demonstration time data were obtained under non-optimal observing
conditions the $B$ photometry, acquired in time-series mode, allowed us to identify 274  
variable stars (among which 96 are bona fide and 31 are candidate RR Lyrae stars, 71 are Cepheids, and 16 are binary systems) 
by applying the image substraction technique to selected 
portions of the observed fields.
%
Differential flux light curves were obtained for the vast majority of these variables.
Our sample includes mainly pulsating stars which populate the instability strip from the 
Classical Cepheids down to the RR Lyrae stars, thus tracing the different stellar generations in these regions of M31
 down to the horizontal branch of 
the oldest ($t \sim 10$ Gyr) component.
\end{abstract}


\section{Introduction}

The Andromeda galaxy, our nearest giant neighbor, is the best
place to study the structure, formation and evolution of a massive
spiral and to get hints on whether the merger/accretion or the cloud collapse
is the dominant mechanism in the formation of a giant spiral. Our
external view of the system is less affected by selection and
line-of-sight effects that make it difficult to perform a homogeneous global
study of the Milky Way (MW). As a result, the structures observed in
M31 are easier to study, thus making Andromeda the best current
laboratory for investigating faint stellar structures and merging
events occurring around large galaxies.

van den Bergh (2000, 2006) suggested that M31 originated as an early
merger of two or more relatively massive metal-rich progenitors. This
would account for the wide range in metallicity (Durrell, Harris, \&
Pritchet 2001, Stephens et al. 2001, Bellazzini et al. 2003) and age (Brown et al. 2003) observed in the M31 halo,
compared with the MW. M31 hosts spectacular signatures of present and
past merging events, such as the giant tidal stream (Ibata et al.  2001)
that extends several degrees from the center of the galaxy (McConnachie
et al. 2003), and the arc-like overdensity connecting the galaxy to
its dwarf elliptical companion NGC 205 (McConnachie et
al. 2004). Metallicity, luminosity, and alignment with M32 are
consistent with the M31 stream being tidally extracted from M32. A number of 
earlier papers following the stream discovery did indeed speculate that M32 might be the source of the stream
 (Ibata e al. 2001, Choi et al. 2002, Ferguson et al. 2002). 
 The low-velocity
dispersion (11 km s$^{-1}$) supports the notion that the giant stream is a coherent
interaction remnant. However, Keck/DEIMOS spectroscopic studies (Font et al. 2006), the models for the 
progenitor and the $N$-body studies of its tidal disruption tend to argue against 
M32 as being the stream progenitor (Fardal et al. 2008). 
The infrared photometry of the M31 disk obtained by IRAS
(Habing et al. 1984) and by the Spitzer Space Telescope (Barmby et
al. 2006) revealed two rings of star formation off-centered from the
M31 nucleus (Block et al. 2006, and references therein).  The two
rings seem to be density waves propagating in the disk. Beccari et
al. (2007) unveiled the near-ultraviolet view of the M31 inner ring
using the blue channel camera of the Large Binocular Telescope (LBT/LBC-blue). 
Numerical simulations show that both rings may result from a
companion galaxy plunging head-on through the center of the M31 disk
about 210 Myr ago (Block et al. 2006).

Ferguson et al. (2002) and Ibata et al. (2007) presented the first
panoramic views of the Andromeda galaxy, based on deep Isaac Newton
Telescope (INT) and Canada-France-Hawaii Telescope (CFHT) photometric
observations that cover, respectively, the galaxy's inner 55 kpc and the
southern quadrant out to about 150 kpc, with an extension that reaches
M33 at a distance of about 200 kpc. 
Their data show the whole giant stream, including its extensions and also reveal 
a multitude of streams, arcs and many other large-scale structures of
low surface brightness, as well as two new M31 dwarf companions (And
XV and And XVI).  Twelve dwarf galaxies were known to be M31
companions until 2004, among which only 6 dSphs and an additional 17 
new M31 satellites were discovered in the last five to six years, mainly based
on the INT and CFHT data.  The most recent census is reported by
Richardson et al. (2011) along with the discovery of five new M31 dSph
satellites, And XXIII-XXVII, as a result from the second year of data
of the ``PanAndromeda Archaeological Survey" (PAndAS; McConnachie et
al. 2009). Started in August 2008, the PAndAS survey is extending the
global view of M31 and of its companion M33 by collecting data with
MegaPrime/MegaCam on the 3.6 m CFHT, for a total area of 300 square
degrees and out to a maximum projected radius from M31's centre of
about 150 kpc. This survey will likely detect several other M31 faint
satellites.

The primary tool for understanding the formation history of a galaxy over the whole Hubble time is the analysis of
color-magnitude diagrams (CMDs) deep enough to reach the main-sequence turnoff (TO) of the oldest
populations. The TO of the oldest stars in M31 ($V \sim $ 28.5 mag) is still unreachable by the largest
ground-based telescopes, and it requires  tens of  orbits of HST/ACS time 
devoted to ``tiny" ($3^{\prime}.5 \times 3^{\prime}.7$) portions of the galaxy
(Brown et al. 2003, 2006, 2007, 2008, and 2009) to be reached and measured. The shallower survey by Richardson et al. (2008), 
who obtained CMDs 
%
to $V \sim $ 27.5 mag for 
14 HST/ACS fields and sampled different substructures around M31, covers a total area which is roughly 1/3 
of that allowed in a single shot by a ground-based telescope 
like the LBT.    

Pulsating variable stars offer a powerful alternative tool to
trace stars of different ages in a galaxy because variables of different
types arise from parent populations of different ages.  Specifically, the RR Lyrae stars and the Population II Cepheids (often referred to as Type II
Cepheids, T2Cs), which belong to the oldest stellar population (t $> 10$ Gyr), 
allow us to trace the star formation history (SFH) back to the
first epochs of galaxy formation and, with their mere presence, can provide crucial
insight on the timescale and merging episodes that may have led to
the assembly of a galaxy.  These low-mass variables ($M \leq \sim 1 M_{\odot}$) burn  
helium in the stellar core and hydrogen in the shell surrounding the core during the
horizontal branch (HB, the RR Lyrae) and post-HB (the T2Cs) evolutionary phases. They are respectively
about 3 and 4 magnitude
brighter, and hence much easier to observe, than 
old (t $>$ 10 Gyr) TO stars. 
%
The typical shape of their light variation, which occurs with periodicities in the range of 0.2 to less than 1 day for the RR Lyrae stars and
from 1 to 25 days for T2Cs, makes them much
easier to recognize, even when the upper HR diagram has
overlapping contributions from a complex mix of age and metallicity.
 The Anomalous Cepheids (ACs) are intermediate-age (t$\sim$ 1-2 Gyr) variables with 
periods in the range of  $\sim$0.3 to $\sim$2.0 days and mean magnitudes spanning a 1.5 mag range.
These variables are intrinsically brighter than the RR Lyrae stars 
 and, when observed in an external stellar system where distance and projection effects can be neglected, they locate themselves along a strip,
  with the 
least luminous ones (at periods $\sim$0.30 days) being  about 0.5 mag brighter than the RR Lyrae stars, and the most 
luminous ones (at periods $\sim$2.0 days) being about 2.0 mag 
brighter than the RR Lyrae stars.  They appear 
significantly brighter than the T2Cs, at fixed period.
 From an evolutionary point of view they are metal-poor ([Fe/H] $\leq -1.7$) 
He-burning stars in the post-turnover portion of the ZAHB (see Marconi, 
Fiorentino, Caputo 2004 and references therein) that cross the 
instability strip (IS) at a higher luminosity than the RR Lyrae variables. Their 
mass is around 1.5 M$_{\odot}$. With an age of a few Gyr,  they indeed represent  the 
natural extension of Classical Cepheids (CCs) to lower metallicities and 
masses (see e.g. Caputo et al. 2004, Fiorentino et al. 2006 and 
references therein). Finally, CCs are intermediate-mass (typically from 3 to 12 M$_{\odot}$) stars that cross the IS 
during the blueward excursion in the 
central He-burning phase, also referred to as a 
blue loop. They have pulsation periods in the range of $\sim$1 to $\sim$100 days and absolute 
visual magnitude of $-2$ to $-7$ mag. With a typical age ranging from a few 
to a few hundred Myr, they represent excellent tracers of the properties of young 
Population I stars.

Short- and intermediate-period variables (i.e., periods shorter than a few days)
of M31 have never been adequately studied despite their great
potential to trace the stellar populations and SFHs of nearby galaxies (Mateo 1998, 2000; Saha 1999, Catelan
2004, Clementini et al. 2004). So far, the main limitations have been the
size of the telescopes employed in the ground-based surveys (Pritchet
\& van den Bergh 1987, Dolphin et al.  2004, Vilardell et al. 2007, Joshi et al. 2010), the limited number of
available HST archive data (Clementini et al. 2001), and the rather 
small areas of M31 covered by the space observations (Brown et
al. 2004, Sarajedini et al. 2009, Jeffery et al. 2011).  In their deep HST/ACS survey of
the M31 halo Brown et al. (2004) identified 55 RR Lyrae stars (29
fundamental-mode $-$RRab$-$, 25 first overtone $-$RRc, and one
double-mode $-$RRd$-$ pulsators) in a $3^{\prime}.5 \times
3^{\prime}.7$ field along the southeast minor axis of the
galaxy. Based on their pulsation properties, Brown et al.  concluded
that the old population in the Andromeda halo has Oosterhoff intermediate properties and 
does not conform to the
subdivision in Oosterhoff I (Oo~I) and Oosterhoff II (Oo~II)
types (Oosterhoff 1939)
 followed by the RR Lyrae stars in the MW field and 
globular clusters (see Clementini 2010 and references therein).  In this respect, the M31 halo field would thus be
different from the MW halo field, and this would point to a different
formation/evolution history. However, a different conclusion was
reached by Sarajedini et al. (2009) who identified 681 RR Lyrae
variables (555 RRab's, and 126 RRc's) based on HST/ACS observations of
two fields near M32, at a projected distance between 4 and 6 kpc from
the center of M31, and concluded that these M31 fields have Oosterhoff
I properties. More recently, Jeffery et al. (2011) have extended Brown et al.'s study of the M31 halo RR Lyrae stars to five additional ACS/HST
fields that include: one pointing in the stream, one pointing in the disk, and three pointings in the M31 halo, at distances of 
21 kpc (hereafter referred to as halo21) and 35 kpc from the galaxy centre (hereafter halo35a and halo35b, respectively; see Fig. 1 
in the Jeffery et al. paper). 
They detected 21 RR Lyrae stars in the ``disk"
field, 24 in the ``stream" field, 3 in the ``halo21" field, 5 in the ``halo35b" field, and none in the ``halo35a" field. 
Field ``halo21'' is of particular interest 
to us since it overlaps with our field H1. Jeffery et al. found average periods for the fundamental mode pulsators of
0.583, 0.560 and 0.495 days in the disk, stream, halo21 and halo35b fields, respectively, and conclude that the RR Lyrae populations in these 
M31 fields appear 
to be mostly of Oosterhoff type I.
But, how general are all of these results, given the small
areas covered by Brown et al. (2004), Sarajedini et al. (2009), and Jeffery et al. (2011) 
studies?  Another open question is whether there are any RR Lyrae
stars in the M31 giant stream and, if any, whether their properties differ from the properties of the variables in the surrounding M31 
fields. Jeffery et al. 
(2011) detected 24 RR Lyrae stars in their ``stream'' field.  These RR Lyrae stars could either belong to
the merged satellite, to M32 where RR Lyrae stars are claimed to exist
 (Alonso-Garcia, Mateo \& Worthey 2004, Fiorentino et al. 2010), or
have formed during the merger, in which case the merging event would
have occurred at least 10 Gyr ago. In any case, their pulsation
properties could provide hints to identify the progenitor stream
stars. According to the average periods in Jeffery et al.'s Table 2, the ``stream'' RR Lyrae stars seem 
to be more OoI-like than the variables in the other fields they observed. However, given the small number statistics and the rather 
limited field of view of ACS/HST, this could simply be a statistical artifact. Clearly, sampling of much larger areas is needed to draw any
general conclusions on the properties of the M31 RR Lyrae population. 

\begin{figure*}
\epsscale{1.9}
\plotone{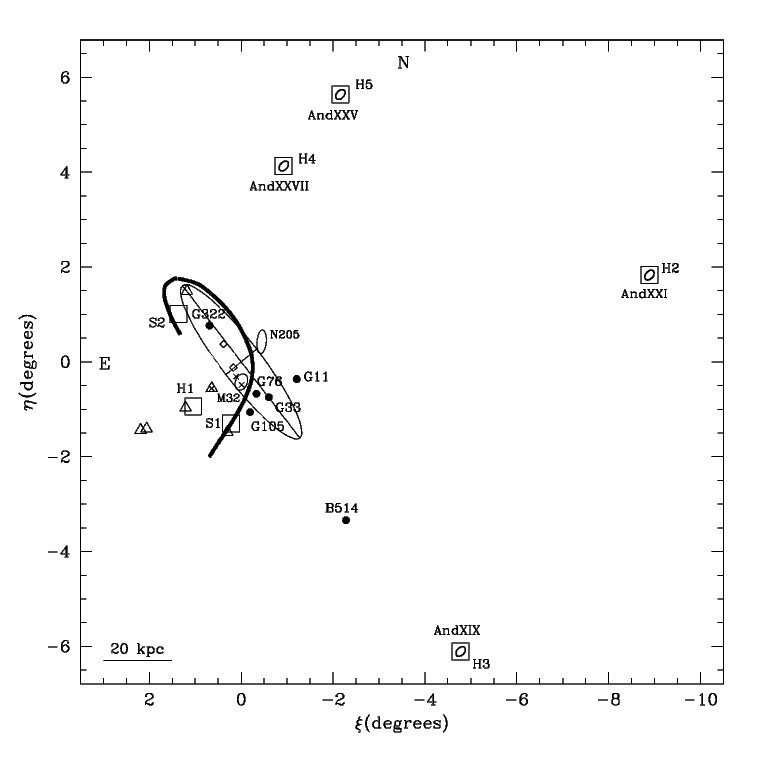}
\caption{Schematic map of the Andromeda galaxy showing the location of the fields targeted with the LBT/LBC-blue (boxes), and the
GCs observed with HST (filled circles). 
 The heavy-solid line shows 
the approximate location of the M31 giant stream according to Ferguson et al. (2002).
Crosses show the centers of the M31 fields studied by Brown et al. (2004) and Sarajedini et al. (2009) using ACS/HST time-series data.
Open triangles show the M31 fields studied with ACS/HST by Jeffery et al. (2011).
Diamonds mark the center positions of the M31 fields studied with ground-based time series observations by Vilardell et al. (2006, 2007) 
and Joshi et al. (2010).
\label{map}}
\end{figure*}

To address the above questions, we are carrying out a long-term project to study the 
stellar populations of both constant and variable stars in properly selected fields and GCs
of the Andromeda galaxy, as well as in recently discovered M31 dSph satellites. We have used the Wide Field Planetary Camera 2 (WFPC2) on board the Hubble Space
Telescope (Cycle 15 HST program GO 11081, PI: G. Clementini) 
to resolve the cluster stars (see Clementini et al. 2009, 
Contreras  2010), and the 
wide field and light-collecting capabilities of the LBT to monitor portions of the M31 giant stream and 
halo, and four among the most extended M31 dSphs: And XIX, And XXI, And XXV, and And XXVII. An ESO Large Program (ID: 186.D-2013, PI G. 
Clementini) is also in progress at the Gran Telescopio Canarias (GTC) to study the variable stars and stellar populations of five further M31 dSphs. 
Figure~\ref{map} shows the location of
the target fields and GCs   
(squares and filled circles, respectively) of our HST and LBT observations on a schematic map of the Andromeda galaxy. 
Also shown in the figure are 
the fields studied for variability by Brown et al. (2004), Sarajedini et al. (2009), and Jeffery et al. (2011) using ACS/HST time-series data, and 
the fields studied for variability by Vilardell et al. (2006, 2007) and Joshi et al. (2010) using ground-based facilities. 

Results for the cluster variables show that the RR Lyrae stars in the M31 GCs may have different properties than their MW
counterparts (Clementini et al. 2009, Contreras Ramos 2010, Contreras Ramos et al. 2011, in preparation). 
The LBT is an ideal tool for studying the pulsating variable stars in the M31 field and in its dSphs because it reaches   
the same level of accuracy as the HST studies, but on a much larger area, thus allowing us to attain a statistical 
significance never reached before for an external galaxy because each LBT field covers an area about 37 times larger than an HST/ACS-WFC field. 

The identification and center coordinates of the 
M31 regions that we are observing with the LBT 
 are provided in Table~\ref{lbt_fields}. The stream fields were  
chosen  to monitor both the stream 
portion that enters into the M31 disk (field S1 in Fig.~\ref{map}) and a region 
toward the northeast portion of the stream that exits from the disk (field S2 in Fig.~\ref{map}). 
Fields H1, H2, H3, H4 and H5 were instead chosen to provide a representation of different portions of Andromeda's 
 halo on the opposite 
sides of the galaxy. 
 In particular, 
fields H2 and H3 are located at about 122 kpc northwest and 104 kpc southwest of the M31 centre,  respectively, and contain two new 
M31 dSphs, And XXI (Martin et al. 2009) in field H2  
and And XIX (McConnachie et al. 2008) in field H3. Fields H4 and H5 are located along a filamental structure 
at about 57 kpc and 81 kpc northest of the M31 centre, respectively, and contain two of the most recently discovered M31 satellites, And XXVII 
and And XXV (Richardson et al. 2011). Finally, field H1 is 
at about 19 kpc from the center of M31 in the southeast direction.

In this paper, which is part of our series on the study of variable stars in M31, we present results from pilot observations of 
fields S2 and H1 with the blue channel of the Large 
Binocular Camera (LBC-blue) 
mounted at the prime focus of the first unit of the LBT (Giallongo et al. 2008) obtained during the LBT Science Demonstration Time (SDT). 
Each of these fields covers 
a 23$^\prime \times 23^\prime$ area.
We have obtained  
$B,V$ CMDs down to $V \sim 26$ mag  for both fields.
 The large field of view along with the high sensitivity of LBT/LBC-blue allowed us 
 to bridge portions of the M31 disk to traces 
of the galaxy giant stream in a single shot of field S2. Similarly, the southwest corner of the halo field H1 probably  includes the 
southeast portion of the giant stream.
We present results of a search for variable stars in these regions of the Andromeda galaxy.
A number of technical problems and rather unfavourable weather/seeing conditions hampered our observing campain. 
Nevertheless, using the image subtraction technique we were able to 
identify and obtain
differential flux light curves for a number of 
CCs with periods in the range of 3 to 10 days, a few candidate ACs and/or, more likely,  short-period CCs (spCCs) 
with periods around 1-2 days, more than 100 RR Lyrae stars, and a number of binary systems, in the portions of fields S2 and H1 
where the image subtraction technique worked out properly. 

Observations, data reduction and calibration of the photometry are discussed in Section 2. The CMDs 
of field S2 and H1 are presented in Section 3. Results on the variable stars and the catalogue of light curves are presented
in Section 4. Finally, a summary and 
discussion of the results are presented in Section 5.  
%
%

\section{Observations and Data Reduction}
\begin{figure*}
\includegraphics[scale=0.60]{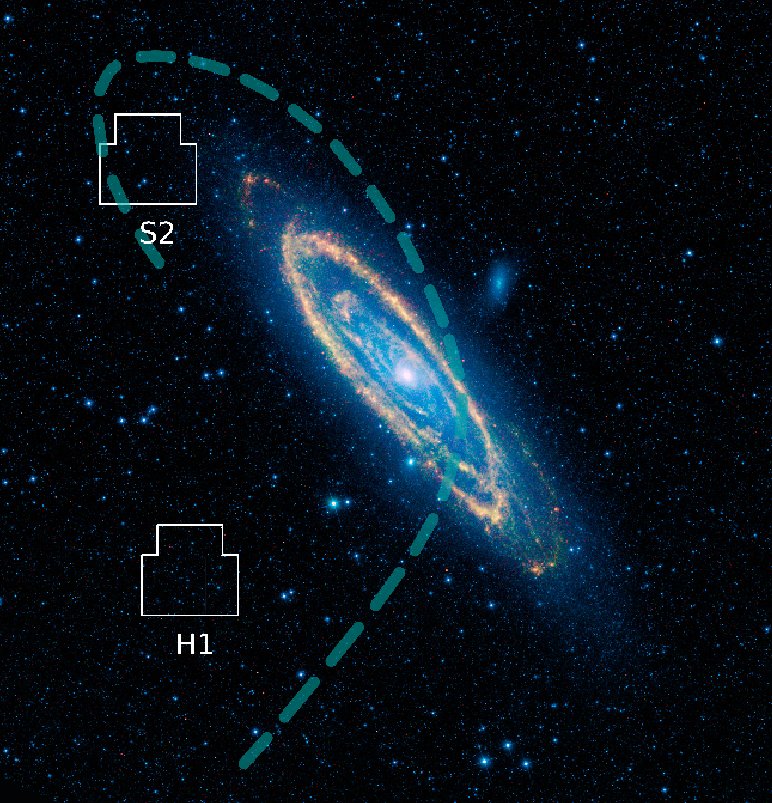}
\caption{$3.5 \times 3.5$ deg$^2$ image of the Andromeda galaxy obtained from the combination of the 3.4 $\mu$, 4.6 $\mu$, 12 $\mu$ and 
22 $\mu$ fluxes, 
measured by the NASA's Wide-field Infrared Survey Explorer (WISE; Image Credit: NASA/JPL-Caltech/UCLA), 
showing the location of 
fields S2 and H1 and a schematic view of the M31 giant tidal stream (heavy dashed line).}
\label{field_S2}
\end{figure*}

$B,V$ photometry of the M31 fields S2 and H1 (see Table~\ref{lbt_fields}) 
was obtained with the LBT/LBC-blue, during ten hours of SDT 
 of the Blue Channel in 2007 October 11-18.  
Given the LBT/LBC-blue scale (0.225 arcsec/pixel) and the total field of view (FOV, 23$^\prime \times 23^\prime$), each of these 
fields covers an area roughly corresponding to 5.1$\times$5.1 kpc$^2$ at the distance of M31 ($\mu_{\rm M31} \sim 24.4$ mag).
Fig.~\ref{field_S2} shows the location
of fields S2 and H1 over a 
$3.5 \times 3.5$ deg$^2$ image of the Andromeda galaxy obtained from the combination of the
3.4 $\mu$, 4.6 $\mu$, 12 $\mu$ and 22 $\mu$ fluxes
measured by the NASA's Wide-field Infrared Survey Explorer (WISE), 
along with a schematic view of the M31 giant tidal stream.
%

Both our fields are contained in the area surveyed by Ferguson et al. (2002) with the INT, reaching a limiting magnitude $V \sim 24.5$ mag,
i.e., 1.5 mag shallower than our photometry. These areas are also planned to  
be observed by PAndAS, with 
 limiting magnitude $g_0 \sim 25.5$ mag; however, no CMDs of the regions sampled by 
fields S2 and H1 have been published yet.
Field H1 contains the two HST/ACS fields of the M31 ``Minor axis" observed by Richardson et al. (2008; see their Table 1), 
and the field ``halo21" (see Fig.~\ref{map}) observed by Jeffery et al. (2011).

The RR Lyrae stars in M31 are expected to have average  magnitudes  around $V \sim$ 25.3-25.5 mag. Taking into 
account their typical intrinsic 
 colors,  amplitudes and periods ($B-V \sim 0.2-0.4$, $A_V \sim$ 0.3-0.5 and 0.6-1.2 mag, P$\sim$ 0.2-1 day, for first 
 overtone and fundamental mode pulsators, respectively) we aimed at reaching a limiting magnitude 
 of $B\sim$26 mag (corresponding to the minimum light of these variables in M31)  in no longer than 15-20 min to avoid smearing the light curve and to 
 have an acceptable 
 S/N even at the light curve minimum. 
 Based on the LBT exposure time calculator, we had estimated that in dark time with a 15 min exposure and seeing=1 arcsec we would obtain 
 S/N$\sim$6 for $B$=26 mag,
  and S/N$\sim$9 for $V$=25.5 mag.
This would have been perfectly adequate for our purposes. Unfortunately, seeing conditions varied significantly
during our observing run, ranging from 0.8 to 2.7 arcsec. We also
experienced a number of problems with the focus and
tracking of the telescope during these early phases of LBT operations, which did not allow us to make individual exposures longer
than 300 sec. Our observations, which were acquired in time-series mode, consist of 59 $B$ and
8 $V$ frames of field S2, and 48 $B$ and 3 $V$ frames of field
H1, each frame corresponding to a 300 sec exposure, and  
we obtained an S/N$\sim$2 for $B \sim$ 26 mag in our best image with
FWHM$\sim$0.8 arcsec. 
Notwithstanding the unfavourable weather and technical conditions, we obtained 30 $B$ and 6 $V$
images of field S2, and 33 $B$ and 1 $V$ image of field H1 with FWHM
$<$ 1.3 arcsec which, allowed us to identify 
a
candidate variable stars as faint as 
$V \sim$ 25.5 mag in the portions of fields S2 and H1 less affected by optical distortions, where we succeeded to apply the 
image subtraction technique (ISIS, Alard 2000). 
It should also be noted that 
the $V$ images of both fields S2 and H1 were accidentally 
trimmed during the readout of the CCDs; as a consequence, the upper 500 pixels of each CCD in the $V$ images were lost. 



Pre-reduction of the entire data set (bias-subtraction and
flat-fielding) through the LBC dedicated pipeline was provided by the
LBC team\footnote{http://lbc.oa-roma.inaf.it/}.  PSF-photometry of the
pre-reduced images of each chip of the LBC mosaic was
then performed with DoPHOT
(Schechter et al. 1993) on the two images obtained in the best observing
conditions (1$B$ and 1$V$ with FWHM$\sim$ 0.8-1 arcsec for each of
the two fields) to produce the CMDs. This package allowed us to model the stellar PSF, which varies significantly along each CCD of our LBC frames, 
much more efficiently  than DAOPHOT. On the other hand, our attempt to use 
DAOPHOTII/ALLSTAR/ALLFRAME
(Stetson 1987,1994) to process the individual time-series data and
produce light curves on a magnitude scale for the variable stars very often failed due to both the geometric distortions and poor FWHM of 
the vast majority of our frames. For this reason we obtained light curves on a magnitude scale only for a very limited number of 
variable stars located in small portions of the frames where DAOPHOTII/ALLSTAR/ALLFRAME was run successfully.
%
%
A 2MASS catalogue\footnote{http://irsa.ipac.caltech.edu/} was used to
identify astrometric standards in the LBC field of view. 
More than a thousand
stars were used to find an astrometric solution for each of the 
LBC CCDs. 
 Accuracy of the derived coordinates is on the order of $\sim$ 0.3-0.4 arcsec (rms) in both
right ascension and declination.
The absolute photometric calibration of the S2 and H1 photometry was
obtained using a set of 192 local secondary standard stars with $B,V$
photometry in the Johnson-Cousins system, extracted from the Massey et
al. (2006) catalogue, and falls in the region of field S2 covered by
CCD 1.
Aperture corrections were separately calculated for each of the 4 CCD mosaics of
fields S2 and H1 by performing aperture photometry in each photometric 
band with the SExtractor package (Bertin et al. 1996). They are provided in Table~\ref{apcor}.  
The derived calibration equations are:
$$B = b - 0.0635(b-v) + 27.78 - K_bX_b$$
and       
$$V = v + 0.0107(b-v) + 28.12 - K_vX_v$$
\noindent
where $B$ and $V$ are the standard magnitudes and $b,v$ are the instrumental
magnitudes normalized to 1 sec and corrected for aperture corrections
using the values given in Table~\ref{apcor}. $K_b$ and K$_v$ are the extinction coefficients
in $B$ and $V$ for which we adopted values of 0.22 and 0.15 mag, respectively, as provided on
the 
LBC commissioning web page (available at http://lbc.oa-roma.inaf.it/commissioning/standards.html). Typical internal errors
of our photometry for non-variable stars at the level of the M31
horizontal branch ($V \sim 25.5$ mag) are $\sigma_{V}$= 0.17 mag, and
$\sigma_{B}$= 0.26 mag, respectively, as provided by the 
DoPHOT reduction of individual $B,V$ images corresponding to 300 sec exposures. 

\section{Color magnitude diagrams}
\begin{figure*}
\epsscale{1.5}
\plotone{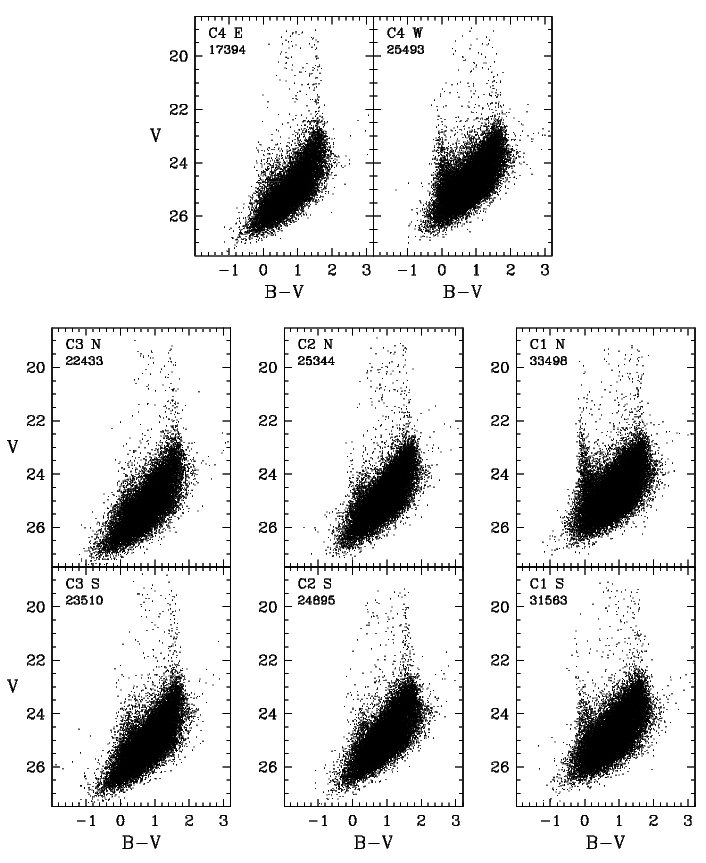}
\caption{$V,B-V$ CMDs of field S2, from a 
pair of $B,V$ images of 300 sec exposure time,
obtained in optimal observing conditions
(FWHM $\sim$ 0.8-1.0 arcsec). In each panel we label the number of stars displayed. 
}
\label{cmd_4chipv_s2}
\end{figure*}

\begin{figure*}
\epsscale{1.5}
\plotone{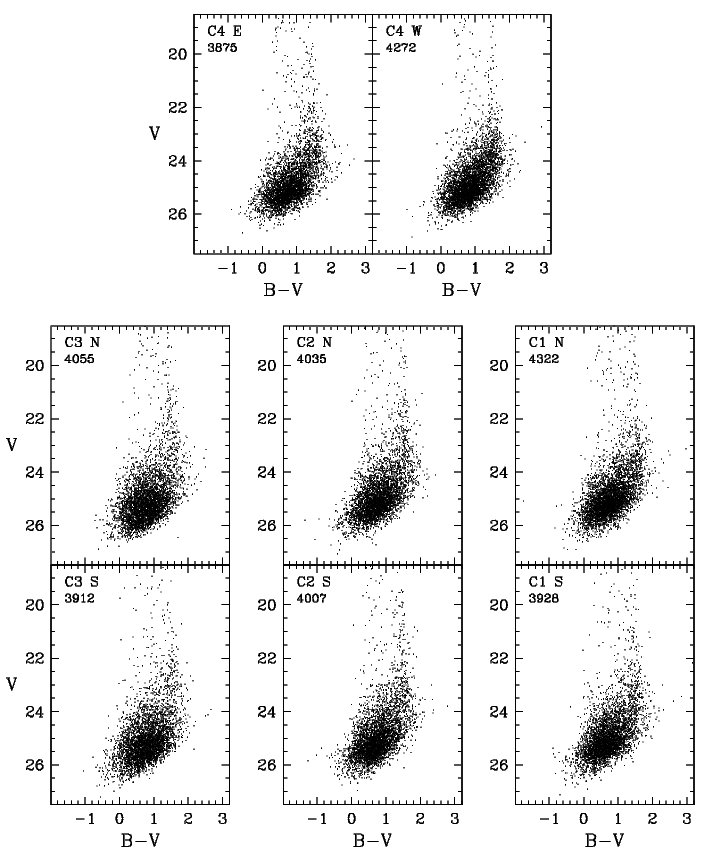}
\caption{$V,B-V$ CMDs of field H1, from a 
pair of $B,V$ images of 300 sec exposure time,
obtained in optimal observing conditions
(FWHM $\sim$ 0.8-1.0 arcsec).
}
\label{cmd_4chipv_h1}
\end{figure*}

Figures \ref{cmd_4chipv_s2}, 
\ref{cmd_4chipv_h1} 
 show the $V,B-V$ CMDs of the 4 CCD 
mosaics of field S2 and H1, respectively, obtained 
at the end of the reduction and calibration processes 
from the DoPHOT photometry of pairs of $B,V$ images of each field, each corresponding to 300 sec exposures obtained with FWHM of about 0.8-1.0 arcsec.
The photometric catalogues producing these CMDs were cleaned from stars with photometric errors larger than twice the mean error 
at each magnitude and by manually removing ``spurious stars" produced by ghosts and spikes of saturated sources and 
 background galaxies. 
%
%
%
%
%
%
%
In each figure, the CMDs are arranged according to the geometry of the 4 CCDs composing the LBC-blue mosaic,  
and each CCD
was divided in 2 equal parts: north and south parts for CCDs 1, 2 and 3, and east and west parts for CCD 4. 
Accordingly, 
 CMDs corresponding to the four different CCDs of each field were labeled as follows: C1 N and C1 S for CCD1 north and south part, respectively, 
and similarly with CCD2 and 3, while the east and west parts of CCD4 were labeled as C4 E and C4 W, respectively.
Each CCD of the LBT/LBC-blue mosaic covers about a 1.7$\times$3.9 kpc$^2$ area of M31; however, because of the trimming of the 
$V$ images, 
the CMDs corresponding to the individual CCDs in fact cover a reduced but still remarkable 
area roughly the size of 1.7$\times$3.4 kpc$^2$.
We have accounted for this problem when dividing CCDs and corresponding CMDs in parts to ensure that each CMD in  
Figures \ref{cmd_4chipv_s2}, 
\ref{cmd_4chipv_h1}, 
 samples the same area of M31. 
The most striking feature in the CMDs of field S2 is a conspicuous 
blue plume observed in panels C1 N, C1 S and  C4 W of Fig. \ref{cmd_4chipv_s2} 
  at $V \leq$ 25.0 mag and
$B-V \leq$ 0.4 mag. This blue plume is barely discernible in C2 N, and eventually disappears, moving eastward from CCD2 to CCD3. 
Also intriguing is a feature seen in C2 N and S, C3 N and S, and C4 E at  
$V \leq$ 25.0 mag and $0.2 < B-V < $ 0.4 mag. 
Finally, all of the CMDs show a variably populated, bright red plume and a sparse distribution of
bright stars of intermediate colors. We believe that the blue plume is produced by young stars possibly associated with an M31 spiral arm and 
the galaxy disk, while the red plume is due to local M dwarfs. 

The CMDs of field H1 (see Fig.\ref{cmd_4chipv_h1}) 
 are much less populated than those of field S2, and the blue plume is 
totally absent, which is not surprising if the blue plume in field S2 is due to the disk and spiral arm stars  
and field H1 is instead representing the M31 halo population. 

\begin{figure*}[ht]
\begin{center}
\includegraphics[scale=0.9]{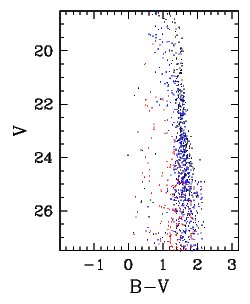}
\caption{Foreground simulation for field S2 including thin disk (black dots), thick disk (blue dots) and halo (red dots)
  stars. This CMD is on the same scale as those of Fig.~3 to allow for a direct comparison.}
\label{cmd_synth}
\end{center}
\end{figure*}

In order to correctly interpret the features we see in the CMDs in terms of the SFH 
 and the structure of M31, a reliable evaluation of the
foreground contamination due to our Galaxy is necessary. To approach
this problem, we have run simulations using a well-tested star-count
code for our Galaxy (see Cignoni et al. 2008, and Castellani et
al. 2002). In this code the MW is divided into three major
Galactic components, namely, the thin disk, the thick disk, and the halo. For
each of these three components an artificial population is created by a random choice of mass
and age from the assumed initial mass function 
 and star formation law, 
  interpolating on a grid of
evolutionary tracks (from the zero age main sequence to the white dwarf phase), the
metallicity of which is determined by the adopted age-metallicity relation. 
Reddening and photometric errors of the data are convolved with magnitudes of  
the synthetic stars, producing a realistic CMD. The
thin disk and the thick disk density laws were modeled by a double
exponential with the same scale length (3500 pc) but with a different scale
height (1 kpc for the thick disk, 300 pc for the thin disk). The halo
follows a power law decay with exponent 3.5 and an axis ratio of
0.8. A local spatial density of 0.11 stars $pc^{-3}$ was adopted for
the thin disk, whereas thick disk and halo normalizations were
$1/10$ and $1/500$, respectively, relative to the thin disk. The
metallicity of each Galactic component was fixed at Z=0.02, Z=0.006 and Z=0.0002 for the 
thin disk, thick disk and halo, respectively. In order to establish quantitative limits to the Galactic star counts
in field S2, all free model parameters were let to vary. In particular, the
thin disk scale height was allowed to vary between 250 and 300 pc, with the 
thick disk and halo normalizations tested between
$1/10$ and $1/20$ and between $1/500$ and $1/850$ relative to the thin
disk. Table~\ref{tab_cignoni} summarizes the predicted star counts as a function of the 
magnitude and color over an area equivalent to the area covered by 
each of the CMDs shown in the 8 panels of Figs.
\ref{cmd_4chipv_s2} and \ref{cmd_4chipv_h1}. Figure \ref{cmd_synth} shows a typical simulated CMD for the foreground
contamination in field S2, which was obtained by assuming $E(B-V)$=0.08 mag, and the typical internal errors 
of our photometry (0.007$< \sigma_B <$0.296 mag and 0.008$< \sigma_V <$0.252 mag, for 20.0$< V <$26.0).
The simulation  describes the contamination 
by Galactic stars, affecting each of the CMDs shown in the 8 panels of Figs.
\ref{cmd_4chipv_s2} and \ref{cmd_4chipv_h1}.
The simulation demonstrates that the Galactic contamination is generally negligible at any magnitude level for
$B-V \leq$ 0.4 mag; hence, the blue plume observed in the CMDs of panels C1 N, C1 S and  C4 W 
is produced by M31 stars, and is not due to contamination by Galactic stars. Conversely, all of the
bright stars with intermediate colors are likely MW stars (of the halo and thick disk), and most of 
the bright red plume stars are MW thick disk M dwarfs.
To make a more quantitative comparison, we have counted the number of stars (as a function of the same 
magnitude and color bins as in  the simulation)  
in  each of the CMDs shown in the 8 panels of Figs. \ref{cmd_4chipv_s2} and \ref{cmd_4chipv_h1}. 
These counts are provided in Tables~\ref{conteggi_s2} and \ref{conteggi_h1} for fields S2 and H1, respectively.
The comparison with Table~\ref{tab_cignoni} shows that the  MW  contamination clearly dominates all the CMDs of
field S2 for magnitudes brighter than $V$=21 mag, both in the blue and the red bins.
In the $21 < V \leq $ 22 mag 
range the MW dominates in the eastern CCDs (CCD4 E and CCD3 N and S) but the M31 contribution increases progressively as we move westward and  
approach the M31 disk and, possibly, a spiral arm.
Similarly, in the $22 < B-V \leq $ 23 mag bin there is an almost equal contribution of MW and M31 stars in the eastern CCDs,
but M31 takes over progressively and becomes dominant in the western CCDs (CCD4 W and CCD1 N and S).
Finally, M31 stars dominate all of the CMDs for magnitudes fainter than $V$=23 mag.
Star counts for field H1 (see Table~\ref{conteggi_h1}) have a smoother distribution, which is expected for a halo population. 
The M31 stars only dominate for magnitudes 
fainter than $V$=23 mag,
while for  $V <$ 23 mag MW and M31 stars contribute almost equally for $0.0 < B-V < $ 0.5 mag,  and the MW generally dominates
for $0.5 \leq B-V < $ 1.0 mag.
%

\begin{figure*}[ht]
\begin{center}
\includegraphics[scale=.72,  clip=true]{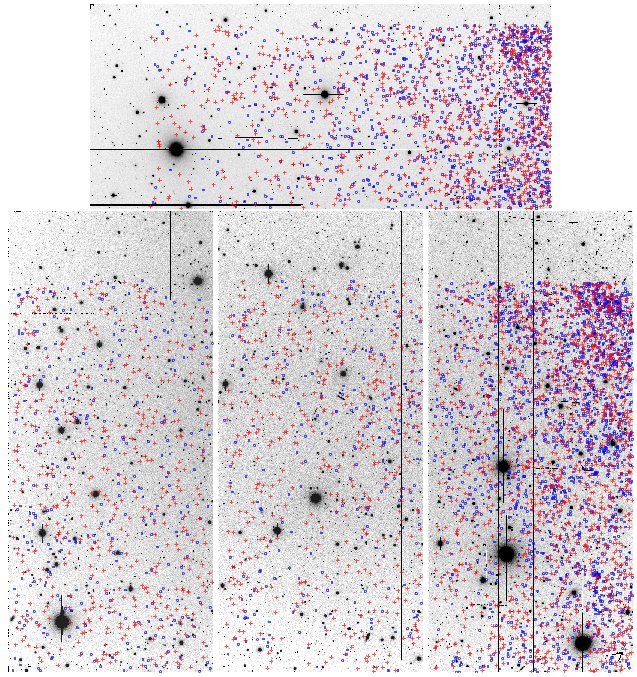}
\caption{Position on the 4 CCDs mosaic of field S2 of stars in the blue plume of the CMD (see Fig.3).
Blue boxes are stars with $V \leq 25.0$ and $B-V \leq 0.2$ mag; red crosses are stars with $V \leq 25.0$ and $0.2 < B-V \leq 0.4$ mag.
}
\label{map_blueplume_S2}
\end{center}
\end{figure*}

\begin{figure*}[ht]
\begin{center}
\includegraphics[scale=.72,  clip=true]{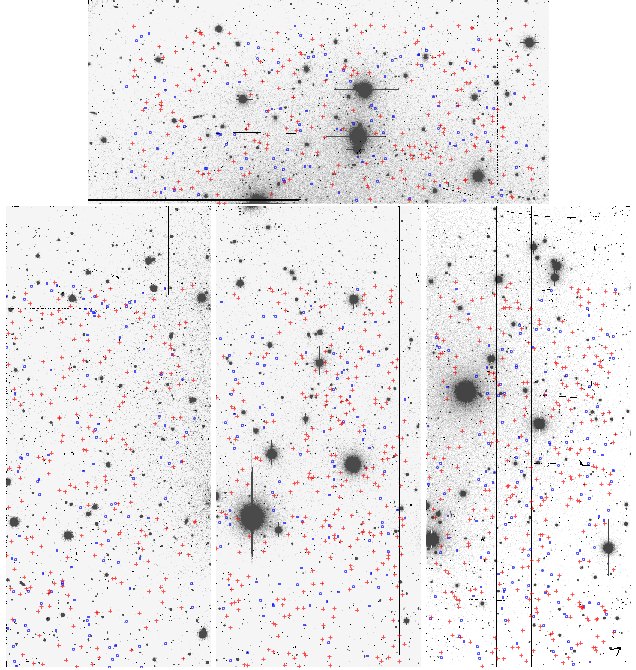}
\caption{Position on the 4 CCDs mosaic of field H1 of stars in the CMD (see Fig. 4) with $V \leq 25.0$ and $B-V \leq 0.2$ mag (blue boxes); and
with $V \leq 25.0$ and $0.2 < B-V \leq 0.4$ mag (red crosses).
}
\label{map_blueplume_H1}
\end{center}
\end{figure*}

In Figs.~\ref{map_blueplume_S2} and \ref{map_blueplume_H1} we show a $B$ image of field S2 and a $B$ image of field H1, respectively, where 
we have overplotted in blue stars with
$ V \leq$ 25.0 mag and
$B-V \leq$ 0.2 mag, which correspond to sources populating the blue plume of the CMDs,
and in red stars having $V \leq$ 25.0 mag and $0.2 < B-V \leq $ 0.4 mag, which correspond to the intermediate-color features seen 
in Figs. \ref{cmd_4chipv_s2} and \ref{cmd_4chipv_h1}. For stars located on the upper 500 pixels of each 
CCD of the mosaic,  we only have $B$ magnitudes, because of the unfortunate trimming of the $V$ images. This is why all of these stars 
are missing in the CMDs of Figs. \ref{cmd_4chipv_s2}, 
 \ref{cmd_4chipv_h1},  
 as well as in the 
images shown in  Figs.~\ref{map_blueplume_S2} and \ref{map_blueplume_H1}.
Nevertheless, while the intermediate-color sources (red crosses) are almost homogeneously spread on all 4 CCDs, both in field S2 and in field H1, 
and thus likely trace the halo component, 
the blue plume stars (blue boxes) appear to be mainly
concentrated 
in the upper-right (northwest) part of CCD1 and in the right (west) portion of CCD4 of field S2, thus likely tracing the disk and, possibly, 
a spiral arm of M31.
To evaluate the significance of these uneven distributions, we have counted the number of stars 
in the blue and intermediate plumes 
of each of the CMDs shown in the 8 panels of Figs.
\ref{cmd_4chipv_s2} and \ref{cmd_4chipv_h1}, respectively, and  in the magnitude bins $ V \leq$ 24.0 mag and $ 24 <V \leq$ 25.0 mag, separately.
  These counts are provided in Tables~\ref{conteggi_s2bis} and \ref{conteggi_h1bis} for
fields S2 and H1, respectively. The star counts in Table~\ref{conteggi_s2bis} show that the number of blue and intermediate-plume sources in field 
S2 increases dramatically, but not homogeneously, as we move westward, from CCD4 E to CCD4 W and from CCD3 to CCD1, and 
approach the M31 disk.  The highest concentration of blue and intermediate-plume stars is found in CCD4 W and CCD1 N, but it drops significantly 
in CCD1 S. The counts in Table~\ref{conteggi_h1bis} instead confirm the smooth stellar distribution in field H1, showing only a marginal
increase in the number of blue and intermediate-plume stars with $24< V \leq 25$ mag in CCD1 N and CCD1 S, where the southwest corner of halo field
H1 perhaps touches a southeast portion of the giant stream (see Fig.~\ref{field_S2}).

\section{Variable Stars}
As anticipated in Section 2, the poor seeing conditions and technical problems 
 made it rather challenging to use our data for the original purpose of studying the variable stars in these regions of M31. A 
crucial complication was the significant optical distortions of the LBT/LBC-blue camera (see Giallongo et al. 2008, Figure 4), 
 particularly 
in the initial operation phase of the LBT.
We had to implement a number of different procedures and conduct several trials to 
detect the variable stars. Therefore, the number of variables we were able to identify is very limited if compared, for instance,  to 
the number one would expect by extrapolating the number densities in Brown et al. (2004) study. However, our fields are much more external 
than Brown et al.'s and, in fact,  our number densities are in much better agreement 
with the number of RR Lyrae stars found by Jeffery et al. (2011) in their ``halo21" field that overlaps with our field H1. This will be reviewed
in further detail in Section 4.5. 
In the following section, we briefly describe the procedures we have implemented and the results we have obtained 
from the search for variable stars in CCD2 and the upper half portion of CCD1 of field S2, and in the upper half 
of CCD2 of field H1. 
\subsection{Identification of the variable stars and light curves}
To identify candidate variables in our $B$ time series images of fields S2 and H1,  we 
used the optimal image subtraction 
technique and the package ISIS2.1 (Alard 2000), which is known to be very efficient at identifying 
variables with amplitudes as low as $\Delta B<$0.1 mag in crowded fields. 
The package was run on the $B$ time series of CCD 1 and 2 of field S2 and CCD 2 of field H1.
We encountered several difficulties 
in aligning and interpolating the images of our LBT/LBC-blue time series data with ISIS, which was likely due to the significant distortions of  
LBT/LBC-blue camera. 
Since the regions of the LBC mosaic less affected by optical distortions are those covered by CCD2, and
the best observing conditions occurred during the observations of field S2, we managed to properly align and interpolate a 
subset of 43 $B$ images of the entire CCD2 of field S2 with ISIS, and then make the subsequent search 
for variable stars. Interpolation did not succeed instead for the entire 
CCD1;  we had to divide it into two halves, and only images corresponding to the upper half of CCD1 of field S2 were successfully aligned.
We encountered even more problems with the images of field H1, since they were generally obtained under worse seeing conditions.
We divided the CCD in two parts and were able to align and interpolate only a subset of 33 images corresponding to the upper half of CCD2.
After aligning and interpolating the images we built reference images of CCD2 - S2, CCD1 - S2 (upper part), and CCD2 - H1 (upper part). We 
subtracted them out from
the respective time series and summed the difference of the images to obtain {\it var.fits} images which, according to ISIS, are the maps of 
variable sources in the frames under study. 
Specifically, we used 19 and 28 frames to build two
{\it var.fits} images of CCD2 of Field H1, 17 and 28 images for CCD2 of field S2, and 20 and 43 images for 
CCD1 of field S2. In order to pick up candidate 
variables from the {\it var.fits} images that were as faint as the RR Lyrae stars, which at minimum light in our frames are expected to
 have an S/N$\sim$ 2, we had to use a very low 
detection threshold of  0.33. We ended up with rather large lists of about 4000 candidate variables from each {\it var.fits} frame. Lists 
corresponding to the pair of {\it var.fits} frames of each field were cross-correlated, thus obtaining about 2000 common candidate sources
per set of images. A careful inspection of these stars returned a final catalogue of 143 bona fide variables in CCD2 of field S2, 
96 variables in the upper 
portion of CCD1 of field S2, and 33 variables in the upper portion of CCD2 of field H1. Two additional 
bona fide variables were also identified in
the lower half of CCD1 of field S2 during a preliminary search with ISIS on the whole CCD1 of field S2. Hence, the total number of variable
stars we were able to identify was 274.

We note that many of the original 
candidate variables could be real variables, but we only retained those that showed periodic, unquestionable, and better sampled light curves.
A summary of the total number of retained candidate variables per field, found with the above procedure, is given in Table~\ref{candvar}. 
Identification (namely ISIS ID, and DoPHOT ID when available), coordinates, and a rough estimate of the period, obtained running 
the Period Determination by Phase Dispersion Minimization (PDM; Stellingwerf 1978) algorithm within IRAF
on the differential $B$ flux time-series of these bona fide candidates is provided in Table~\ref{candvar_coord}. We note that only a very small
fraction of the candidates in
Tables ~\ref{candvar} and \ref{candvar_coord}
have a counterpart with reliable photometry in the ALLFRAME catalogues, and hence, have a light curve on 
a magnitude scale, while the vast majority 
only have $B$-band differential flux light curves.
A number of different problems caused the ALLFRAME PSF photometry of the individual phase-points 
of the variables to be generally unreliable.  
These problems included crowding, particularly in the disk field (field S2); rather poor and varying seeing  
during the observations; and technical problems with the focus and tracking of the telescope, which made the 
FWHM vary strongly along the frames. All of these different effects combined together so that PSF photometry could be obtained only
in a few cases often only for the pair of frames at 0.8 arcsec 
FWHM. The faintest variables were generally detected only with the image subtraction, and 
no ``reliable" PSF photometry could be obtained for most of them with ALLFRAME; on the other hand,  the  
brighter variables had poorly sampled light curves due to the longer periods. 
Even in the halo field (field H1), where variables were also searched using the Stetson 
variability index on the catalogues produced by the ALLFRAME reductions of CCD2, visual inspection of 
the images of many candidates showed that they often had extended PSFs  caused by spikes, CCD 
defects, telescope tracking problems, and, in turn, unreliable photometry. 
In conclusion, while the present data allowed us to identify variable stars, follow up photometry 
in better technical/seeing conditions will be needed to produce light curves on a magnitude scale and 
fully characterize these variables. However, publishing the identification and differential flux light curves obtained in the present study  
will help future variability studies in these regions of M31.

%
%
%

\begin{figure*}
\begin{center}
\includegraphics[scale=0.70]{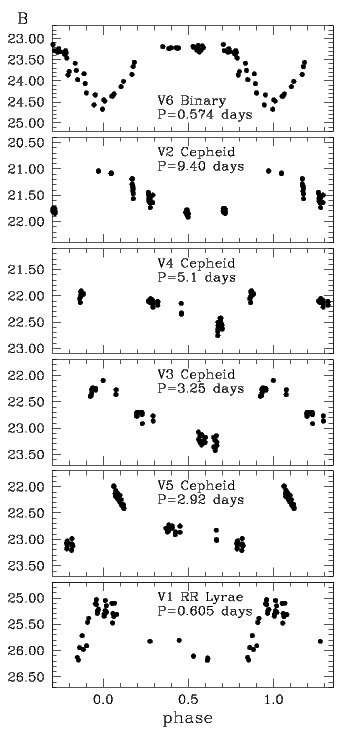}
\caption{Examples of $B$ light curves for 4 Cepheids, an RR Lyrae star, and a binary system detected in CCD1 of field S2. 
Each data point corresponds to a 300 sec exposure. Typical error bars of the
individual data points are in the range from 0.01 to 0.02 mag for the Classical
Cepheids, 0.11-0.17 mag for the candidate Anomalous/short period Cepheids,  and from 0.13 to 0.38 mag 
for the
RR Lyrae stars.
}
\label{light_curves}
\end{center}
\end{figure*}

The study of the light curves of a few of the bona fide candidate variables with light curve 
on a magnitude scale was performed with the
Graphical Analyzer of TIme Series (GRATIS), which custom software developed at the Bologna
Observatory by P. Montegriffo, (see, e.g., Di Fabrizio 1999, Clementini 
 et al. 2000).  
 In Figure~\ref{light_curves} we show examples
 of the $B$ 
 light curves of some variables in field S2 for which we have light curves on a magnitude scale and a reasonably complete coverage of the 
 light cycle.
 They include four  pulsating stars with periods of 9.4, 5.1, 3.25 and 2.92 
days that we have classified as CCs on the basis of their 
brightness and position in the CMD (see below), an RR Lyrae star with 
period of 0.605 days, and a binary system with period of 0.574 days. 
The identification and properties of these six variables are provided in Table~\ref{lbt_var}. Unfortunately, the PSF photometry was not good enough to 
obtain light curves on a magnitude scale for any of the candidate
ACs/spCCs with periods around 1 day. 
$B$-band differential flux light curves for all candidate variables that we were able to identify 
we have identified  
 are presented in Figs.~\ref{light_curves_s2_chip1_p1},~\ref{light_curves_s2_chip2_p1}, and ~\ref{light_curves_h1_chip2_p1}, 
which are published in their entirety 
in the online journal.

\subsection{Classification of the candidate variables}
Since we only have differential flux light curves for the vast majority of the candidate variables in Table~\ref{candvar_coord}  
we do not have information on their magnitude and on the amplitude of their light variation. This complicates the identification 
of the type of variability, since 
the only characteristic parameters we can use to classify the variables are the preliminary period and the shape of the light curve. 

The candidate variables have periods in the range from 0.12 to 9.4 days. Thus, although our observing
strategy was mainly devised to optimize the detection of RR Lyrae stars, it also turned out to be adequate to identify longer period variables. 
 According to the range in period spanned by the candidate variables,  our sample
is likely to contain: RR Lyrae stars (0.2$<$ P $<$1 days), Anomalous (0.3$<$ P $<$2.5 days) and Population II (P $<$ 10 days) Cepheids, 
and short and intermediate period CCs (1$<$ P $<$10 days).
For 138 candidate variables we also have an indication of magnitude because they were measured on the
pair of $B,V$ images of field S2 and H1 with FWHM$\sim$ 0.8 arcsec and thus have $B,V$ magnitudes from the 
DoPHOT photometry (see Table~\ref{candvar}).
Although the DoPHOT magnitudes for the variables correspond to values at random phase on the light curves, they allow us to place the candidates 
on the CMDs 
(see Figs.~\ref{cmd2ter}, ~\ref{cmd2ter_s2chip2}, and ~\ref{cmd2ter_h1chip2}) and thus give us some  
hints about the type of variability.
The location on the CMDs and the periodicities of the variables at $V \sim$ 25-25.4 mag confirms they likely are  
 RR Lyrae stars tracing the
HB of the M31 old stellar component and, perhaps, Population II Cepheids (although the tentative periods generally below 
1 day make a P2C classification unlikely), while 
variables having $V \leq$ 24 mag are likely short- and intermediate-period CCs. 
On the other hand, the classification  of the candidates located more than one magnitude above the
HB at $V$ in the range of 23.5 to 24.5 mag is not easy since the luminosity would suggest they are ACs, while the periods, generally well below 
one day, would make them more likely RR Lyrae stars. However, the AC hypothesis does not seem consistent with the typical metal abundance of the 
stellar population in these M31 fields, but, if these candidates are RR Lyrae stars, their brightness appears to be inconsistent (i.e., too bright) 
with the luminosity of  stars at the red giant branch tip, unless these variables are contaminated (i.e, blended) by other stars. In this 
respect, it is interesting that 
no such intermediate luminosity candidates were detected 
in field H1, which is definitely less crowded that field S2. This point will be discussed in more detail in Section 4.4.
To classify the candidate variable stars we have combined the information on the period, shape of the light curve, and position on the
 CMD (when available). 
We also visually inspected the $B,V$ images with FWHM$\sim$ 0.8 arcsec at the position 
of each candidate variable detected by ISIS, thus revealing the saturated sources, CCD defects, and other problems 
(see notes of Table~\ref{candvar_coord}), as well as 
objects too faint to be reliably measured with DoPHOT,  which could still be tentatively classified. 
The shape of the light curve also revealed several eclipsing binary
systems (see Figs. ~\ref{light_curves_s2_chip1_p1},~\ref{light_curves_s2_chip2_p1},~\ref{light_curves_h1_chip2_p1}) among which a
number of detached systems are certainly worthy of further investigation. 
The variability types deduced from this procedure are provided in Column 8 of Table~\ref{candvar_coord}, where  
uncertain periods or type classifications have been flagged with a question mark.
Our sample includes 96 bona fide and 31 candidate RR Lyraes, 54 bona fide and 17 candidate Cepheids (classical, anomalous or short
period), 14 bona fide and 2 candidate binary systems.  
For the remaining 60 variables no unambiguous classification was possible. However, 
 the unclassified objects are likely to include a number of main sequence variables (see e.g., Baldacci et al. 2005) such as 
$\beta$ Cepheids (P $<$0.3 days) and Be stars (0.4 $<$ P $<$ 3 days) populating the blue-plume at $B-V \sim 0.0$ mag.

Figs.~\ref{cmd2ter}, ~\ref{cmd2ter_s2chip2}, and ~\ref{cmd2ter_h1chip2}
 show the CMDs of the upper part of CCD1 of field S2, the whole CCD2 of field S2, and the upper part of CCD2 of field H1 respectively.
  The candidate variables are plotted as large filled circles, and we have used different colors for the different types of 
variability. 
In the figures, the long-dashed lines around $V$ =25.2 mag  show the boundaries of the
theoretical IS for RR Lyrae stars (Di Criscienzo, Marconi, Caputo 2004), and those around $V$ =24.5 mag the boundaries of 
the IS of ACs with
Z=0.0004 and 1.3 $<$ M$<$ 2.2 M$_{\odot}$ 
(Marconi et al. 2004). This is the highest metallicity allowed for ACs\footnote{As reviewed by
Caputo (1998) for low-metal abundances (Z$\le 0.0004$) and
relatively young ages ($\le$4 Gyr, the effective temperature of ZAHB models
reaches a minimum ($\log{T_{\rm e}}\sim3.76$) for a mass of about 1.0-1.2
$M_{\odot }$, while if the mass increases above this value, both the
luminosity and the effective temperature start increasing, forming the so
called ``ZAHB turnover'' from which ACs are expected to
evolve. For larger metallicities, the more massive ZAHB structures have
brighter luminosities but effective temperatures rather close to the
minimum effective temperature, so that ACs are not
predicted. 
Observationally, ACs are mainly detected in the very metal poor dwarf
spheroidal galaxies and rarely in GCs.}. 
The dotted heavy lines instead represent the first overtone and fundamental blue edges (blue lines) and the fundamental 
red edge (red line) for CC models with Z=0.008, Y=0.25 and 3.25 $<$ M/M$_\odot <$ 11 (Bono, Marconi, Stellingwerf 1999; 
Bono et al. 2002). 
To plot the theoretical IS boundaries on the CMDs we have adopted 
$E(B-V)$=0.08 mag, which was obtained by interpolating on the Schlegel et al. (1998) maps, A$_{V}$=3.315 $E(B-V)$ and A$_{B}$=4.315 $E(B-V)$ 
from Schlegel et al. (1998), and $\mu_0$(M31)= 24.43 mag. 
The latter value was obtained by correcting 
the distance modulus measured by McConnachie et al. (2005) from the M31 red giant branch tip for $E(B-V)$=0.06 mag and  
A$_{I}$=1.94 $E(B-V)$ (Schlegel et al. 1998) to our adopted reddening of $E(B-V)$=0.08 mag.

 It should be noted that these variables are plotted in the CMDs  
using magnitudes and colors sampling random phases of the $B$ and $V$ light curves because we generally have for the variables  few 
measurements of magnitude, and in many cases, we only have the pair of $B,V$ magnitudes that correspond to the two best images used to build the CMDs.  
They span a very large range in color and fall well beyond 
the boundaries of the ISs because of the decoupling 
of their $B$,$V$ magnitudes. 
The six variables listed in Table~\ref{lbt_var} have a better sampling of their light curves in magnitude scale, and hence they are  
also plotted in Fig.~\ref{cmd2ter} using the  $\langle V \rangle$ magnitudes and 
$\langle B \rangle$ - $\langle V \rangle$ colors we obtained by averaging over the full light cycle (large open stars), and with 
arrows connecting the average values to the random phase DoPHOT magnitudes. When the average values are used,  
 the Cepheids and RR Lyrae stars populate 
the rather narrow region of the CMD that corresponds to the classical IS. 
 This exercise illustrates  
how much the candidate variables plotted at random phase could move in the $V, B-V$ plane, and it also demonstrates the 
detection potential and analysis capabilities of our study, which is the main purpose
of the present paper. 
%
%

We also note that, although the number of $B$ phase points would, in principle, be adequate to
obtain a reliable estimate of the $B$ amplitude and 
average magnitude of the variable stars, coverage of the $V$-band light curve is very sparse (only 3-8 phase points in 
the best cases; see Table~1). 
To recover the average magnitudes in the visual band, and thus be able to plot correctly the six variable stars on the CMDs, 
we used the star's $B$ band light curve as
a template, and we properly scaled it in amplitude to fit the few available $V$ data-points. To constrain the scaling factor for the RR Lyrae stars 
we used amplitude ratios $A(B)/A(V)$ computed 
  using literature $B,V$ light curves of
RR Lyrae stars with good light curve parameters. 
These were selected from a number of Galactic globular clusters (GGCs; see Di Criscienzo et al. 2011, for details) and we used amplitude ratios and
phase lags taken from Freedman (1988) and Wisniewski \& Johnson (1968) for the Cepheids. Finally, we used an amplitude ratio of 1
for the binary system, since we do not expect to see significant chromatic effects for binaries.

Given that the RR Lyrae stars are  at the faint limit of our photometry (V$\sim$ 25.5 mag), where the completeness
is dropping off rapidly, understanding the completeness of our study would be important.
Our data clearly suffer from a number of shortcomings that hamp a traditional analysis, making it difficult to estimate the fraction
of variables in each class.   
Fortunately, our field H1 overlaps with Jeffery et al. (2011) field ``halo21", thus allowing a direct comparison of the 
number of RR Lyrae stars found by the two studies and allowing us to draw some conclusions on the completeness of our variable star detections.
Jeffery et al. (2011) found 3 RR Lyrae stars in their 3.5$^\prime$$\times$3.7$^\prime$ ACS/HST field ``halo21", which corresponds to a number density 
of 0.23 RR Lyrae/arcmin$^2$. According to Table~9, we have found 15  bona fide  RR Lyrae stars in the 7.2 $\times$ 8.6 arcmin$^2$ upper 
portion of field H1 
we analyzed for variability, corresponding to a number density of 0.24 RR Lyrae/arcmin$^2$. 
The two numbers compare very favorably, thus showing that, on the assumption that 
the number density of RR Lyrae stars does not vary significantly through field H1, we have attained a good completeness 
of RR Lyrae detections in this field. 

However, our field S2 is too far away from Jeffery et al.'s ``disk" field to make a similar comparison meaningful.

\subsection{Spatial distribution of the variable stars}
Out of the total sample of 274 bona fide candidate variables identified in our study, we have a
magnitude estimate for 138 stars. They include 83 objects with firm classification and 55 objects of uncertain
types.   Among the latter are 8 variables with the typical magnitudes of ACs that will be discussed more 
in detail in Section 4.4.
The subdivision in types of these 138 stars is summarized in Table~\ref{var_type}.
In order to check whether the spatial distribution of the different types of variables can provide some hints on
the underlying parent stellar population, we have also subdivided the variables into the various subfields where they are located.


Discarding the lower half of CCD1 of field S2 where the identification of variables with the image subtraction failed, and with the
caveat that our statistics can, in general, be incomplete, it is interesting to note that the number of RR Lyrae stars remains
almost 
constant in the 4 subfields we have analyzed with ISIS, while the number of CCs decreases dramatically in field H1 where only 2 CCs were detected.
This is consistent with the RR Lyrae stars tracing the M31 halo and with field H1 being a typical halo field of M31. 
Further, the number of CCs also drops significantly as we move from north (upper half of CCD1) to southeast (lower half of 
CCD2), moving away from 
the disk of M31. The highest concentration of CCs is found in the upper part of CCD1 of field S2, suggesting that this 
region may be crossed by a
spiral arm of M31 that also produces the blue plume in the CMD and the candidate main sequence variables (namely, Be and $\beta$ Cepheid stars).  
Finally, the 8 supposed ACs all are found in field S2, while field H1 does not seem to contain candidate variables at $V \sim$ 24 mag. 

\subsection{Stellar populations of different age and chemical composition}
The differences in stellar population shown by the CMDs of fields S2 and H1
are consistent with the different types of variable stars found in these regions of M31 and, as noted in Section 4.3, 
 confirm that H1 
is a typical halo field. The small number of variables detected in CCD2 of field H1 and the lack of variables in
the range of $V \sim$ 23 to 24 mag are consistent with the rather
peripheral location and low stellar density of this field. 
%

Similarly, both the appearance of the CMD and the properties of the variables in the 
 various portions of field S2 sampled, 
respectively, by the whole CCD2 and the upper part of CCD1 reveal significant differences and confirm the  complexity of the 
stellar population 
in this
region of M31.  
With reference to Figs.~\ref{cmd2ter} and ~\ref{cmd2ter_s2chip2}, the portion of field S2 imaged on CCD2 not only lacks the CMD blue plume  
and the
main sequence bright variables, but it also has a few candidate variables at intermediate luminosity ($V\sim$ 22-24 mag). Specifically,
in the whole CCD2 of field S2, only 6 candidate variables are found in the magnitude range
from $V\sim$ 24 to $V\sim$ 22 mag, while 19 such variables are present in the upper portion of CCD1, corresponding to only one-half of the area 
of field S2  
covered by CCD2. Furthermore, almost all of 
the bright variables in CCD2 of field S2 have $V \leq$ 21.5 mag. A large fraction of them is found within
the classical IS, while
the bright candidate variables in CCD1 of field S2 fall, on average, outside of the strip with their random phase magnitudes.  
We note that the three variables marked by orange circles in Fig. ~\ref{cmd2ter_s2chip2}, which could either be ACs/spCCs or bright RR Lyrae 
stars,  are all located in the 
lower part of CCD2.
%
%
%

To get hints on the age and metal abundance of the composite stellar population in field S2, we have compared the CMD 
of the upper part of CCD1 of field S2 (Fig.~\ref{cmd2ter}) with the  
isochrones of Girardi et al. (2002) 
for ages in the range of 63 to 708 Myr and metal abundances of Z=0.008 (Fig.~\ref{cmd2ter_008_gir2002_new}) and Z=0.019 
(Fig.~\ref{cmd2ter_019_gir2002_new}), respectively, to fit the young stellar component in the CMD. To account for the
various types of variables in each figure 
we have also overplotted the theoretical ISs of RR Lyrae stars (Di Criscienzo, Marconi, Caputo 2004), 
ACs with Z=0.0004 and 1.3 $<$ M$<$ 2.2 M$\odot$, and CCs, respectively, with Z=0.008 and Z=0.019. We have then used the mean ridge lines 
of the GGCs M15, M3, M5, NGC2808, 47 Tuc, and NGC6553 (drawn from the $B,V$ database of GGCs by 
Piotto et al. 2002, from Ferraro et al. 1997 for M3, and from Ortolani et al. 1995 for NGC6553) to account for the oldest (t$>$ 10 Gyr) 
stellar component.  
These clusters span a range in metallicity from [Fe/H]=$-0.16$ for NGC6553 to  [Fe/H]=$-2.33$ for M15 on the Carretta et al. (2009, heafter C09) 
metallicity scale. The observed GGC ridge lines were preferred to theoretical isochrones of corresponding age and metal abundance, since the latter 
are known to be affected by large uncertanties in the color transformations from the theoretical to the observational plane.  
To plot ISs, GC ridge lines, and isochrones on the observed CMDs we have assumed:
$E(B-V)$=0.08 mag, A$_{V}$=3.315 $E(B-V)$ and A$_{B}$=4.315 $E(B-V)$, and $\mu_0$(M31)= 24.43 mag. 
%
%
%
%

Figs.~\ref{cmd2ter_008_gir2002_new} and ~\ref{cmd2ter_019_gir2002_new} demonstrate how powerful the approach is of combining information from 
the isochrone and GC ridge line fitting of the CMD with information obtained from the variable star population present in this region of M31. 
In fact, 
the presence of a stellar 
population as old as t$\geq$10 Gyr cannot be unquestionably proven by the comparison of the CMD with the GC ridge lines because of
(a) the confusion of 
stars with different (younger) ages and metal content in the red giant branch region of the CMD  as well as some contamination by Galactic red stars, 
and (b) because  
the HB of such an old population, if any, cannot be disentangled from the overwhelming young population. However, the mere presence of RR 
Lyrae variables having 
an average magnitude consistent with their membership to M31 definitely proves that such an old population exists 
and unambiguously traces the HB of such an old population in this region of M31. 
The properties of the variables also provide hints to the metal abundance of stars in these regions of M31. 
In particular, the average magnitude of the RR Lyrae star having full coverage of the light curve (blue star in Fig.~\ref{cmd2ter_008_gir2002_new}) 
appears to be consistent with the HB of the GCs
M5 ([Fe/H]=$-1.33$, C09) and NGC2808 ([Fe/H]=$-1.18$, C09). The red giant branches of these two clusters also fit rather well the upper envelope 
of the M31 red giant stars,  showing that the bulk of the old (t$\geq$10 Gyr) stellar population in this M31 field has metallicity  
[Fe/H]$ \geq -1.2/-1.3$ on C09 scale.

As for the younger populations, the CMD isochrone fitting seems to favor a 
Z=0.019 metal abundance. However, the position on the CMD of the variables of
a Cepheid type and the comparison with both the isochrones and edges of the theoretical IS for the different metal abundances shows that 
a Z=0.008 component is also needed to explain all the bright variables observed in this field. In fact, although the 9.4 day Cepheid 
(brightest magenta star) and  the  
candidates with $V\leq 21.3$ mag (magenta filled  circles in
Figs.~\ref{cmd2ter_008_gir2002_new}  and ~\ref{cmd2ter_019_gir2002_new}) 
fall on, or close to, the blue loops of the 63 and 79 Myr isochrones  for both the Z=0.008 and Z=0.19,
only for Z=0.008 do  the blue loops extend blueward enough to produce confirmed and candidate CCs with $21.3 \leq V\leq 22.5$ mag.
Furthermore, the position of the four confirmed CCs with respect to the boundaries of the IS suggests
 also  a 
lower metal abundance 
of Z=0.008, as
for $Z=0.019$, they all lie close to the blue edge of the IS. Particularly at 
this metallicity, the two bluest Cepheids are located between the blue edge of the fundamental mode and the blue edge of the first overtone mode. 
This circumstance suggests that if Z=0.019, then the two bluest Cepheids are first overtone pulsators, whereas the other two Cepheids are at the 
fundamental blue edge. Cepheids at the first overtone blue edge are expected to have low pulsation amplitudes. At the same time, the brightest 
Cepheid is at a luminosity for which only the fundamental mode of pulsation is efficient, so that its proximity to the blue edge again implies 
a quite low pulsation amplitude (see e.g. Bono, Castellani, Marconi 2000). These predictions are not consistent with the observed pulsation 
amplitudes that range between 0.84 to 1.29 mag in the $B$ band; they are in better agreement with the values expected if $Z=0.008$ when the Cepheids 
are in the middle of the IS.

Finally, as anticipated in section 4.2, the interpretation of the candidate about 1-1.5 mag above the HB (the orange filled circles in
Figs. ~\ref{cmd2ter}, ~\ref{cmd2ter_s2chip2} and Figure~\ref{cmd2ter_008_gir2002_new}) 
is not easy. Taking into account the magnitude range spanned by these stars ($V$ in the range from 23.5 to 24.5 mag) and the
periods that are generally shorter than 1 day, with only a couple of exceptions, we have considered three different possibilities:
(1) they could be spCCs falling on the short-period tail  (P $<$ 2 days) of the CCs distribution; 
(2) they could be ACs tracing an intermediate-age population as metal poor as [Fe/H] $\leq -1.7$; or 
(3) they could be overluminous RR Lyrae stars.
Altough the uncertainty in the periods and the lack of complete light curves for these objects makes discerning among the three hypotheses
rather difficult, the comparison of the  Girardi et al. isochrones shows 
that  the blue loops of the 
Z=0.008 isochrones do not extend blueward enough 
to cross the IS (see isochrones for  316, 562 and 708 Myr in Fig.~\ref{cmd2ter_008_gir2002_new}) at the low mass and luminosity of these variables,
 thus ruling out possibility number 1,
at least for populations with Z=0.008. In Fig.~\ref{cmd2ter_001_gir2002_new} we show the effect of further reducing the metal abundance 
to Z=0.001 and Z=0.0004, respectively. The blue loop of a 708-Myr isochrone with Z=0.001 can produce these faint variables at least in part,  
as well as the brighter CCs, and would even better 
produced them with Z=0.0004. On the other hand, these variables would also be throughly consistent with the IS 
of the ACs for Z=0.0004. However, if the bulk of the oldest stellar population in these regions of M31 has a metal abundance of 
[Fe/H]= $-1.2$/$-1.3$, as suggested by the RR
Lyrae stars and the fitting with the GC HB ridge lines, it does not seem 
conceivable that the intermediate-age and younger populations in the same field could have metal abundances as low as Z=0.0004 
(corresponding to [Fe/H]= $-1.7$). In
conclusion, this seems to rule out both the case of the ACs (namely, hypothesis number 2) and, within  hypothesis number 1, the case of CCs 
as metal poor as Z=0.0004.
It is also worth noting that the predicted gap in magnitude between the RR Lyrae level and the faintest Cepheid pulsators at 
a metallicity of ~0.004 is of about 0.8 mag (Caputo et al. 2004) in excellent agreement with the difference in magnitude between the RR Lyrae and 
the two AC/spCC pulsators we see in Figs.~\ref{cmd2ter_008_gir2002_new}, ~\ref{cmd2ter_019_gir2002_new} and ~\ref{cmd2ter_001_gir2002_new}. 

Finally, hypothesis number 3, which states that a large fraction of these variables might be RR Lyrae stars, is supported by their periods being generally 
below 1 day.
However, since these variables also appear to be about 1 mag brighter than the HB of the old population in M31, they are either 
 blended with other sources if they are RR Lyrae stars or they cannot belong to Andromeda 
but perhaps to a structure (a satellite or the stream) in the galaxy foreground located in the region covered by field S2. 

It is obviously too premature to draw any firm conclusions based on only a few variables and the rather incomplete light curves available to us.  
Still, these results show how much could be learned on the stellar populations and structure of M31 by combining the study of the galaxy CMD and the 
variable star properties based on LBT data.

\subsection{Comparison with previous variability studies in M31} 
Recent studies of variable stars in M31 include the works of Vilardell, Jordi \& Ribas (2007), and Joshi et al. (2010) for the Cepheids, and the
studies of Brown et al. (2004), Sarajedini et al (2009), and Jeffery et al. (2011) for the RR Lyrae stars.  
Both the Brown et al. (2004) and Sarajedini et al. (2009) fields have different size, are rather far away from our LBT pointings and, 
generally, are much closer to the M31 disk. These aspects, along with the uncertainty of the completeness of the detection of variable stars in 
our LBT fields, make the comparison quite difficult. 
This is particularly true for the RR Lyrae stars since we cannot easily evaluate the completeness of our samples at such faint magnitude 
levels and, on the other hand, our fields are very peripheral compared with those of 
Brown et al. (2004) and Sarajedini et al (2009).
Indeed, our two fields sample regions of M31  at the projected distances of 21 kpc (field S2)  and 19 kpc (field H1) from the center 
of the galaxy, respectively, while  Brown et al. (2004) observed an ACS/HST field at 11 kpc,
 and Sarajedini et al. (2009) observed two ACS/HST fields at 4 and 6 kpc, respectively.
 On the other hand,  one of the Jeffery et al. (2011) fields, field ``halo21", overlaps with our
field H1. The comparison of the number density of RR Lyrae stars in fields H1 and ``halo21" shows that, in spite of all the shortcomings affecting 
our data, we seem to have reached a good completeness in the detection of these variables. 

The comparison is easier for bright variables such as the Cepheids. 
%
Vilardell et al (2006, 2007) detected 416 Cepheids in a 33.8$^{\prime} \times 33.8^{\prime}$ field located along the minor axis of M31 
(see  Fig.~\ref{map}) 
using the 2.5m Isaac Newton Telescope (INT) in La Palma, Spain. This corresponds to an average density of 0.36 Cepheids~arcmin$^{-2}$.
The Cepheids in Vilardell et al. sample have a period distribution roughly peaking around 4 days (see Fig. 2 in their paper), and 
they span the magnitude range of $V\sim 23$  to 
$V\sim 19.2$ mag, which overlaps well with the range in magnitude spanned by Cepheids in our LBT fields. 
Of our LBT pointings, the region imaged in the upper portion of CCD1 of field S2 is the closest to the M31 disk.  
In 
this region we have identified 18 bona fide CCs and another 6 variables with a more uncertain classification, all having a visual 
magnitude in the range of 23 to 19 mag. 
Accounting for the trimming of the $V$ frames, these variables span an area of about 61.44 arcmin$^2$, providing a density of 
0.39 Cepheids~arcmin$^{-2}$, in good agreement with the density findings of Vilardell et al.
Unfortunately, we only have a very preliminary estimate of the periods of our Cepheids; thus, it is not possible to make a sound 
comparison of the period 
distributions. As for 
%
the metal abundances, Vilardell et al. (2007) assume the galactocentric metallicity gradient by Zaritsky et al. (1994) and an empirical 
metallicity correction of the Cepheid Period-Luminosity (PL) relation. According 
to their resulting corrections and to the reported galactocentric distances, we estimate that the Cepheids in their sample have [O/H] in the 
range $-$0.2 $\div$ 0.2. Since we are exploring regions of M31 that are quite external (see Fig.~\ref{map}, and Column 7 of Table~\ref{lbt_fields}),
 we 
expect that the metallicity of our Cepheids is close to 
[O/H]=$-$0.2, which implies $Z\sim 0.01$. This is consistent with the metallicities Z=$0.019-0.008$ we infer for  CCs brighter than V$\sim$ 23 mag  
as compared with the isochrones of Girardi et al. (2002) discussed in Section 4.4.  

As for the comparison with Joshi et al. (2010), those authors have used a 1m telescope and identified 39 short-period  (P$<$ 15 days) CCs 
in a 13$^{\prime} \times 13^{\prime}$ region  of the M31 disk located along the semimajor axis on the same side of our field S2 
(see  Fig.~\ref{map}). This corresponds to a density of 0.23 Cepheids~arcmin$^{-2}$, which is about one-half of the value 
we derive in the
upper part of CCD1 of field S2. On the other hand, 
the Joshi et al. sample contains Cepheids that are generally  brighter and of a longer period than in our sample. Their period 
distribution peaks in fact 
at $\log P \sim$ 0.9 and 1.1 days, with periods 
as short as 3.4 days, and most of their Cepheids are at  $\langle R \rangle \sim$ 20-21 mag.
Our sample includes Cepheids of shorter period, but our magnitudes and periods are consistent with their results. 
Particularly,  we find $\langle V \rangle$=20.62 mag for the 9.4 days Cepheid (V2, see Table~10) and in the same period range they find 
consistent $R$ magnitudes, based on a typical 
$V-R$ color and taking into account the uncertainties related to reddening corrections.

To derive a distance estimate for the four CCs listed in Table~\ref{lbt_var} and to confirm that they are M31 members, we have used the 
theoretical PL and the Wesenheit relations for 
Z=0.008 and Z=0.02 (see Caputo, Marconi, \& Musella 2000) and the PL 
relation adopted by the HST Key Project (Freedman et al. 2001), which has the slope by  Udalski et al. (1999) and the zero point based on an 
assumed distance modulus for the LMC of 18.5 mag.  The
resulting individual (from the Wesenheit relations) and mean (from the
PL) distances are reported in Table~\ref{distance} where $\mu_{PL}$, $\mu_{Wes}$, $\mu_{OGLE_{HST}}$, $\mu_{PL02}$, and 
$\mu_{Wes02}$ are distance moduli derived from the theoretical PL and Wesenheit 
relations for Z=0.008, the PL relation by Freedman et al. (2001), and 
the theoretical PL and Wesenheit relations for Z=0.02. 
We consider an uncertainty of $\pm$ 0.1 mag to individual distance moduli from the
Wesenheit relations, while the average distance modulus obtained 
from the Udalski et al. (1999) PL is 24.57$\pm$ 0.2 mag. 
%
The errors on the estimated distances take into account both the observational
errors in $B$ and $V$ and the intrinsic dispersion of the adopted
relations. Our value is longer, but it is within the errors 
consistent with the modulus of McConnachie et al. (2005) transformed to our adopted reddening of $E(B-V)$=0.08 mag, $\mu_0$(M31)= 24.43 mag.

Vilardell et al (2007) estimated 
a distance to M31 of (m$-$M)$_0$=24.32 $\pm$0.12 mag, based on an assumed distance modulus of 18.4 mag for the Large Magellanic Cloud (LMC). 
The difference in the adopted distance modulus for the LMC combined with the adopted metallicity
 correction explains most of the discrepancy between their distance estimate and our results.

\section{Summary and Conclusions}
We have presented $V, B-V$ CMDs reaching the limiting magnitude $V\sim 26$ mag  of two
fields of the Andromeda galaxy  which were observed with the LBT/LBC-blue camera during the SDT. 
A number of technical problems during the first phase of the LBT/LBC operation and rather unfavorable weather/seeing conditions
 hampered our observing campaign, thereby limiting our study of the variable star populations in these M31 fields.
 Nevertheless, we have identified 274 variable stars using the
image subtraction technique and present their differential flux $B$ light curves.
For 138 of these variable stars we have also obtained an estimate of the $B, V$ magnitudes that allowed us to plot the variables on the CMDs.
By combining information gathered on the period, magnitude, shape of the light curve, and position on the CMD, we were able to classify 
214 variables. They include 127 RR Lyrae stars, 71 short- and intermediate-period Cepheids
(periods shorter than 9.4 days), and 16 binary systems. 
We have compared the CMD and the variable star population of the M31 field closest to the galaxy disk and the giant stream 
with the sets of isochrones by Girardi et al. (2002) 
for ages in the range of 63-708 Myr, and metal abundances of Z=0.0004, 0.001, 0.008,  and 0.019 to fit the young stellar component,  
as well as with the mean ridge lines 
of GGCs spanning a range in metallicity from [Fe/H]=$-0.16$ dex to  [Fe/H]=$-2.33$ dex (on the C09 metallicity scale)
to account for the oldest (t$>$ 10 Gyr)  stellar component.
The isochrone and GC ridge line fittings and the properties of the variable stars show that the composite stellar population present in this M31 
region has typical metal abudance 
larger than [Fe/H]=$-$1.2/$-$1.3 dex for the oldest stellar component and for the young stellar component is in the range 
of $\sim -$1.3 dex to about solar metallicity.
Far from being complete and exaustive,  this study nevertheless demostrates the  powerful approach of combining information from 
the CMD with information obtained from the variable star population.




\bigskip

\acknowledgments

The LBT is an
international collaboration among institutions in the United States, Italy and
Germany. LBT Corporation partners are: The University of Arizona on behalf of
the Arizona university system; Istituto Nazionale di Astrofisica, Italy; LBT
Beteiligungsgesellschaft, Germany, representing the Max-Planck Society, the
Astrophysical Institute Potsdam, and Heidelberg University; the Ohio State
University, and The Research Corporation, on behalf of The University of Notre
Dame, University of Minnesota and University of Virginia.

This publication makes use of data products from the Two Micron All Sky Survey, 
which is a joint project of the University of Massachusetts and the Infrared 
Processing and Analysis Center/California Institute of Technology, funded by 
the National Aeronautics and Space Administration and the National Science 
Foundation.
We thank Paolo Montegriffo for the development and maintenance of the GRATIS package.
This research was partially supported by PRIN-MIUR-2007JJC53X, 
PI F.Matteucci, and by COFIS ASI-INAF I/016/07/0.

\clearpage
 
  \begin{table*}
  \small
      \caption[]{Identification and coordinates of our LBT fields in M31.}
         \label{lbt_fields}
     $$
         \begin{array}{llllrrrcc}
	    \hline
            \hline
           \noalign{\smallskip}
           {\rm Name} & {\rm Object~type}~~~~  & ~~~~~~~~    {\rm \alpha } & ~~~~~~~~    {\rm \delta} & {\rm \xi}~~~~~ & {\rm \eta}~~~~~ & 
	   {\rm D_{M31~center}} &  {\rm N_B}~~ & {\rm N_V}~~ \\  
            ~~        &                        &~~~~  {\rm (2000)}   & ~~~~  {\rm (2000)}& {\rm (degrees)} & {\rm (degrees)} & {\rm (kpc)~~~~~} & {\rm (a)} & {\rm (a)}\\ 
            \noalign{\smallskip}
            \hline
            \noalign{\smallskip}
	    
{\rm H1}  &  {\rm Halo~field          }  & 00~48~13.11& +40~19~09.4 & 1.04~~~&-0.94~~~& 18.9~~~~~ &48	    &3\\
{\rm H2}  &  {\rm And~XXI             }  & 23~54~47.71& +42~28~15.0 &-8.89~~~& 1.83~~~&121.8~~~~~ &59       &56\\
{\rm H3}  &  {\rm And~XIX             }  & 00~19~32.1 & +35~02~37.1 &-4.78~~~&-6.11~~~&104.2~~~~~ &48       &46\\
{\rm H4}  &  {\rm And~XXVII           }  & 00~37~27.1 & +45~23~13   &-0.93~~~& 4.13~~~& 56.9~~~~~ &{\rm (b)}      &{\rm (b)}\\
{\rm H5}  &  {\rm And~XXV             }  & 00~30~08.9 & +46~51~07   &-2.16~~~& 5.64~~~& 81.1~~~~~ &{\rm (b)}      &{\rm (b)}\\
{\rm S1}  &  {\rm Stream~field        }  & 00~43~51.51& +39~58~09.4 & 0.21~~~&-1.30~~~& 17.7~~~~~ &2  & $\nodata$      \\
{\rm S2}  &  {\rm Stream~field        }  & 00~49~08.31& +42~16~09.4 & 1.18~~~& 1.01~~~& 20.9~~~~~ &59		 &8\\
\hline
            \end{array}
	    $$
{\small $^{\mathrm{a}}$ Each image corresponds to a 300 sec exposure for fields S2, H1 and S1, and to 420 exposures for fields H2, H3, H4, and H5.}
{\small $^{\mathrm{b}}$ Observations scheduled for fall 2011.}
\par\noindent

\end{table*}

\begin{table*}
\small
\caption[]{Aperture corrections for the 4 CCD mosaic image of field S2 (upper part) and H1 (lower part).}
\label{apcor}
     $$
         \begin{array}{ccc}
	    \hline
            \hline
           \noalign{\smallskip}
           & {\rm Field~~S2}& \\
            \hline
            \noalign{\smallskip}
&B&V\\ 
{\rm CCD~1}&-0.236& -0.301\\ 
{\rm CCD~2}&-0.251& -0.304\\ 
{\rm CCD~3}&-0.244& -0.254\\ 
{\rm CCD~4}&-0.229& -0.216\\ 
\hline
            \noalign{\smallskip}
           & {\rm Field~~H1}& \\
            \hline
            \noalign{\smallskip}
&B&V\\ 
{\rm CCD~1}&-0.216& -0.134\\ 
{\rm CCD~2}&-0.211& -0.168\\ 
{\rm CCD~3}&-0.222& -0.127\\ 
{\rm CCD~4}&-0.232& -0.109\\ 
\hline
\end{array}
$$
{\small Note. Corrections correspond to aperture minus PSF magnitudes.}
\par\noindent
\end{table*}

\begin{table*}
\small
\caption{Range of expected Galactic contaminating stars as a function of magnitude
  and color over an area equivalent to the area covered by 
each of the CMDs shown in the 8 panels of Figs.
\ref{cmd_4chipv_s2} and \ref{cmd_4chipv_h1}. Blue: 0$<B-V<$0.5 mag; Red: 0.5$\leq B-V<$1.0 mag. }
\label{tab_cignoni}
     $$
         \begin{array}{ccc}
	    \hline
            \hline
&~~~{\rm Blue}~~~~~&{\rm Red}\\
            \hline
18.5 \leq V\leq 20 &0-1&14-24\\  
20<V \leq 21&0-1& 6-10\\ 
21<V \leq 22&0-1& 3-9\\ 
22<V \leq 23&0-1& 3-7\\ 
23<V \leq 24&1-3& 3-8\\ 
\hline
\end{array}
$$
\end{table*}

\begin{table*}
\small
\caption{Number of stars in the four CCD mosaic of field S2 as a function of magnitude and color
over an area equivalent to the area covered by 
each of the CMDs shown in the 8 panels of Fig.
\ref{cmd_4chipv_s2}. Blue: 0$<B-V<$0.5 mag; Red: 0.5$\leq B-V<$1.0 mag.}
\label{conteggi_s2}
     $$
         \begin{array}{cccccccccccc}
	    \hline
            \hline

&&\multicolumn{2}{c}{\rm CCD4~E} &&&&& \multicolumn{2}{c}{\rm CCD4~W} \\
      &&{\rm Blue}&{\rm Red}&&&&&{\rm Blue}&{\rm Red}\\
            \hline
18.5 \leq V \leq 20   &&~2&~16&&&&&~~4&~~5\\
20<V \leq 21&&~3&~12&&&&&~~1&~11\\
21<V \leq 22&&~2&~~8&&&&&~10&~14\\
22<V \leq 23&&~4&~13&&&&&~23&~40\\
23<V \leq 24&&46&148&&&&&165&443\\
\hline
&&\multicolumn{2}{c}{\rm CCD3~N} &&\multicolumn{2}{c}{\rm CCD2~N} & \multicolumn{3}{c}{\rm CCD1~N} \\
      &&{\rm Blue}&{\rm Red}&&{\rm Blue}&{\rm Red}&&{\rm Blue}&{\rm Red}\\
            \hline
18.5 \leq V \leq 20   &&~1&~12&&~~0&~~8&&~~0&~~8\\
20<V \leq 21&&~2&~~7&&~~4&~17&&~~3&~~8\\
21<V \leq 22&&~1&~~4&&~~2&~15&&~15&~~7\\
22<V \leq 23&&~3&~13&&~~4&~23&&~17&~36\\
23<V \leq 24&&35&133&&~38&151&&260&667\\
\hline
&&\multicolumn{2}{c}{\rm CCD3~S} &&\multicolumn{2}{c}{\rm CCD2~S} & \multicolumn{3}{c}{\rm CCD1~S} \\
      &&{\rm Blue}&{\rm Red}&&{\rm Blue}&{\rm Red}&&{\rm Blue}&{\rm Red}\\
            \hline
18.5 \leq V \leq 20   &&~1&~15&&~~2&~~4&&~~2&~13\\
20<V \leq 21&&~1&~~7&&~~6&~10&&~~2&~~8\\
21<V \leq 22&&~2&~~8&&~~3&~~7&&~~7&~11\\
22<V \leq 23&&~7&~14&&~~3&~19&&~~7&~27\\
23<V \leq 24&&65&166&&~36&136&&107&554\\
\hline
\end{array}
$$
\end{table*}

\begin{table*}
\small
\caption{Number of stars in the four CCD mosaic of field H1 as a function of magnitude and color over an area equivalent to the area covered by 
each of the CMDs shown in the 8 panels of Fig. 
\ref{cmd_4chipv_h1}.
Blue: 0$<B-V<$0.5 mag; Red: 0.5$\leq B-V<$1.0 mag.}
\label{conteggi_h1}
     $$
         \begin{array}{cccccccccccc}
	    \hline
            \hline

&&\multicolumn{2}{c}{\rm CCD4~E} &&&&& \multicolumn{2}{c}{\rm CCD4~W} \\
      &&{\rm Blue}&{\rm Red}&&&&&{\rm Blue}&{\rm Red}\\
            \hline
18.5 \leq V \leq 20   &&11&~10&&&&&~~5&~11\\
20<V \leq 21&&~1&~13&&&&&~~3&~~6\\
21<V \leq 22&&~4&~~9&&&&&~~1&~~8\\
22<V \leq 23&&~3&~~9&&&&&~~6&~16\\
23<V \leq 24&&21&~72&&&&&~15&~94\\
\hline
&&\multicolumn{2}{c}{\rm CCD3~N} &&\multicolumn{2}{c}{\rm CCD2~N} & \multicolumn{3}{c}{\rm CCD1~N} \\
      &&{\rm Blue}&{\rm Red}&&{\rm Blue}&{\rm Red}&&{\rm Blue}&{\rm Red}\\
            \hline
18.5 \leq V \leq 20   &&~1&~16&&~~3&~10&&~~7&~~8\\
20<V \leq 21&&~3&~~3&&~~4&~~6&&~~3&~13\\
21<V \leq 22&&~4&~~5&&~~4&~~7&&~~0&~~7\\
22<V \leq 23&&~3&~10&&~~6&~16&&~~6&~10\\
23<V \leq 24&&25&~82&&~25&107&&~22&~83\\
\hline
&&\multicolumn{2}{c}{\rm CCD3~S} &&\multicolumn{2}{c}{\rm CCD2~S} & \multicolumn{3}{c}{\rm CCD1~S} \\
      &&{\rm Blue}&{\rm Red}&&{\rm Blue}&{\rm Red}&&{\rm Blue}&{\rm Red}\\
            \hline
18.5 \leq V \leq 20   &&~4&~~9&&~~5&~12&&~1&~~6\\
20<V \leq 21&&~2&~~5&&~~2&~~7&&~2&~~9\\
21<V \leq 22&&~3&~11&&~~1&~11&&~2&~~5\\
22<V \leq 23&&~5&~10&&~~6&~10&&~6&~12\\
23<V \leq 24&&14&~94&&~31&101&&33&100\\
\hline
\end{array}
$$
\end{table*}

\begin{table*}
\small
\caption{Number of stars in the four CCD mosaic of field S2 as a function of magnitude and color over an area equivalent to the area covered by 
each of the CMDs shown in the 8 panels of Fig. 
\ref{cmd_4chipv_s2}.
Blue plume (BP): $B-V \leq$0.2 mag; Intermediate plume (IP): 0.2$<B-V \leq$0.4 mag.}
\label{conteggi_s2bis}
     $$
         \begin{array}{cccccccccccc}
	    \hline
            \hline

&&\multicolumn{2}{c}{\rm CCD4~E} &&&&& \multicolumn{2}{c}{\rm CCD4~W} \\
      &&{\rm BP}&{\rm IP}&&&&&{\rm BP}&{\rm IP}\\
            \hline
V\leq 24       &&~32&~19&&&&&199&~65\\
24<V\leq 25&&139&195&&&&&679&~649\\
\hline
&&\multicolumn{2}{c}{\rm CCD3~N} &&\multicolumn{2}{c}{\rm CCD2~N} & \multicolumn{3}{c}{\rm CCD1~N} \\
      &&{\rm BP}&{\rm IP}&&{\rm BP}&{\rm IP}&&{\rm BP}&{\rm IP}\\
            \hline
V\leq 24       &&~14&~18&&~18&~27&&344&~82\\
24<V\leq 25&&116&175&&129&220&&740&718\\
\hline
&&\multicolumn{2}{c}{\rm CCD3~S} &&\multicolumn{2}{c}{\rm CCD2~S} & \multicolumn{3}{c}{\rm CCD1~S} \\
      &&{\rm BP}&{\rm IP}&&{\rm BP}&{\rm IP}&&{\rm BP}&{\rm IP}\\
            \hline
V\leq 24        &&~27&~32&&~19&~22&&131&~36\\
24<V \leq 25&&142&220&&~83&166&&469&544\\
\hline
\end{array}
$$
\end{table*}

\begin{table*}
\small
\caption{Number of stars in the four CCD mosaic of field H1 as a function of magnitude and color over an area equivalent to the area covered by 
each of the CMDs shown in the 8 panels of Fig. 
\ref{cmd_4chipv_h1}.
Blue plume (BP): $B-V \leq$0.2 mag; Intermediate plume (IP): 0.2$<B-V \leq$0.4 mag.}
\label{conteggi_h1bis}
     $$
         \begin{array}{cccccccccccc}
	    \hline
            \hline

&&\multicolumn{2}{c}{\rm CCD4~E} &&&&& \multicolumn{2}{c}{\rm CCD4~W} \\
      &&{\rm BP}&{\rm IP}&&&&&{\rm BP}&{\rm IP}\\
            \hline
V\leq 24       &&~8&14&&&&&~8&11\\
24<V\leq 25&&53&87&&&&&45&93\\
\hline
&&\multicolumn{2}{c}{\rm CCD3~N} &&\multicolumn{2}{c}{\rm CCD2~N} & \multicolumn{3}{c}{\rm CCD1~N} \\
      &&{\rm BP}&{\rm IP}&&{\rm BP}&{\rm IP}&&{\rm BP}&{\rm IP}\\
            \hline
V\leq 24       &&~6&16&&~5&19&&~4&~17\\
24<V\leq 25&&46&82&&45&93&&57&109\\
\hline
&&\multicolumn{2}{c}{\rm CCD3~S} &&\multicolumn{2}{c}{\rm CCD2~S} & \multicolumn{3}{c}{\rm CCD1~S} \\
      &&{\rm BP}&{\rm IP}&&{\rm BP}&{\rm IP}&&{\rm BP}&{\rm IP}\\
            \hline
V\leq 24        &&~9&13&&~3&~24&&17&~17\\
24<V \leq 25&&47&81&&49&116&&65&124\\
\hline
\end{array}
$$
\end{table*}

\begin{table*}
\small
\caption[]{Number of bona fide candidate variables identified in field S2 and H1 using the image subtraction technique, and 
numbers of candidates recovered in the ALLFRAME and DoPHOT catalogues.}
\label{candvar}
     $$
         \begin{array}{lcccc}
	    \hline
            \hline
           \noalign{\smallskip}
           \multicolumn{5}{c}{\rm Field~S2}\\
            \hline
            \noalign{\smallskip}
&{\rm N(frames)}&{\rm N(candidates)}&{\rm N(ALLFRAME)} &{\rm N(DoPHOT)}\\ 
{\rm CCD~1~(upper~half)}&43&~~96 & 6 & 49\\ 
{\rm CCD~1~(lower~half)}&$\nodata$&~~~2&0&2\\ 
{\rm CCD~2~(total)}     &43&143&40&74\\ 
\hline
            \noalign{\smallskip}
           \multicolumn{5}{c}{\rm Field~H1}\\
            \hline
            \noalign{\smallskip}
&{\rm N(frames)}&{\rm N(candidates)}\\ 
{\rm CCD~2~(upper~half)}&33&33&$\nodata$&13\\ 
\hline
\end{array}
$$
\end{table*}

\begin{table*}
 \footnotesize
      \caption[]{Identification and characteristics of candidate variable stars identified in the M31 fields S2 and H1}
         \label{candvar_coord}
     $$
         \begin{array}{ccccccclc}
	    \hline
            \hline
           \noalign{\smallskip}
           \multicolumn{9}{c}{\rm CCD1~-~Field~S2}\\
           {\rm Id}    & {\rm Id}      &{\rm \alpha } & {\rm \delta} & {\rm P} & $B$& $V$& {\rm Type}& {\rm Notes}\\
           {\rm (ISIS)}& {\rm (DoPHOT)}&{\rm (2000)}  & {\rm (2000)} & {\rm (days)} & {\rm (mag)} & {\rm (mag)} & &\\
                           &       {\rm (a)}   &              &              &              & {\rm (b)}   & {\rm (b)}   & &\\
            \noalign{\smallskip}
            \hline
            \noalign{\smallskip}

    2783 &$\nodata$ & 00~ 48~ 45.0 &+42~ 21~ 05 & 0.26 &$\nodata$  &$\nodata$  &{\rm RR?}  & {\rm (c)}\\
    2833 &~1576     & 00~ 48~ 44.6 &+42~ 19~ 44 & 0.57 &23.64  &23.64 & {\rm Bin.?}  & \\
           \hline
            \end{array}
	    $$
{\small Table~\ref{candvar_coord} is published in its entirety in the electronic edition of the {\it Journal}, only a portion is shown here 
for guidance regarding its form and content.}\\
{\small $^{\mathrm{(a)}}$ Id(DoPHOT) is the star identification number on the $B,V$ images with FWHM $\sim$ 0.8-1.0 arcsec, 
that were reduced with the DoPHOT package producing all the CMDs shown in the paper.}\\
{\small $^{\mathrm{(b)}}$ $B,V$ values are from the DoPHOT photometry of the $B,V$ images obtained with FWHM $\sim$ 0.8-1.0 arcsec.
 They correspond to values taken at random phase on the light curves. For 6 
variables which have full coverage of the light curve (see Table~\ref{lbt_var}) we list the average values over the full light cycle, while  
random phase values are given in parentheses.}\\
{\small $^{\mathrm{(c)}}$ This candidate variable falls in the portion of the $V$ frame that was trimmed during the read-out of the
CCDs.}\\
{\small $^{\mathrm{(d)}}$ The star is on or close to a dead column of the CCD in the $B$ frame.}\\
{\small $^{\mathrm{(e)}}$ The star  was not measured on the 0.8 arcsec FWHM $V$ frame because too bright and close to saturation.}\\
{\small $^{\mathrm{(f)}}$ Variable star which has full coverage of the light curve, (see Table~\ref{lbt_var}).}\\
{\small $^{\mathrm{(g)}}$ The star is close to the spike of a saturated star.}\\
{\small $^{\mathrm{(h)}}$ The star is close to a dead column of the CCD in the $V$ frame.}\\
{\small $^{\mathrm{(i)}}$ Candidate variable stars identified during a preliminary search with ISIS on the whole CCD1 of field S2.
Stars with DoPHOT IDs 14532 and 5707 are located in the Southern part of CCD1 of Field S2.}\\
{\small $^{\mathrm{(l)}}$ The star is on a defect of the CCD in the $B$ frames.}\\
{\small $^{\mathrm{(m)}}$ The star is contaminated by a bright companion.}\\
{\small $^{\mathrm{(n)}}$ The star is saturated in both the $V$ and $B$ 0.8 arcsec FWHM frames.}\\
{\small $^{\mathrm{(o)}}$ The classification as an AC is mainly based on the star luminosity, but 
is inconsistent with the typical metal abundance of the underlying stellar population (see discussion in Section 4.4).}\\
\end{table*}

  \begin{table*}
 \footnotesize
\tablenum{9}
      \caption[]{continued}
     $$
         \begin{array}{ccccccclc}
            \hline
           \noalign{\smallskip}

    2860 &$\nodata$ & 00~ 48~ 44.1 &+42~ 14~ 52 & 0.52 &$\nodata$  &$\nodata$  &{\rm RRab?}& {\rm (c)}\\
    2883 &$\nodata$ & 00~ 48~ 44.4 &+42~ 20~ 39 & 0.25 &$\nodata$  & $\nodata$ &{\rm RR? } & {\rm (c)}\\
    3310 &~~248     & 00~ 48~ 42.8 &+42~ 15~ 36 & 0.34 &21.02  &19.70  & {\rm CC}  & \\
    3343 &$\nodata$ & 00~ 48~ 43.0 &+42~ 20~ 46 & 0.63 &$\nodata$  & $\nodata$  &{\rm AC?/spCC?/RR?}& {\rm (c,o)}\\
    3425 &11954     & 00~ 48~ 42.4 &+42~ 16~ 02 & 0.63 &24.83  &24.55  &{\rm RR?}  & \\
    3431 &$\nodata$ & 00~ 48~ 42.3 &+42~ 16~ 13 & 0.49 &$\nodata$ &$\nodata$  &{\rm ?}  & \\
    3435 &~~456     & 00~ 48~ 42.5 &+42~ 18~ 36 & 0.52 &22.14  &22.12  &{\rm CC?/Bin.?/Be?}  & \\
    3487 &~~351     & 00~ 48~ 42.2 &+42~ 17~ 07 & 0.64 &21.67  &21.79  &{\rm CC?/Bin.?/Be?}  & \\
    3712 &$\nodata$ & 00~ 48~ 41.7 &+42~ 21~ 44 & 0.32 &$\nodata$  &$\nodata$  &{\rm ?}  & {\rm (c,d)}\\
    3864 &22635     & 00~ 48~ 41.1 &+42~ 19~ 38 & 0.38 &25.17  &25.37  &{\rm RRc}  & \\
    3898 &10857     & 00~ 48~ 41.0 &+42~ 19~ 44 & 0.62 &24.70  &23.11  &{\rm AC?/spCC?/RR?}  &{\rm (o)} \\
    3974 &3330      & 00~ 48~ 40.6 &+42~ 19~ 21 & 1.74 &24.18  &23.92  &{\rm AC/spCC}     &{\rm (o)} \\
    4139 &10043     & 00~ 48~ 39.8 &+42~ 16~ 29 & 0.76 &24.81  &25.37  &{\rm RR}  & \\
    4185 &$\nodata$ & 00~ 48~ 39.8 &+42~ 19~ 24 & 1.46 &$\nodata$  &$\nodata$  & {\rm CC} & {\rm (e)}\\
    4229 &$\nodata$ & 00~ 48~ 39.7 &+42~ 21~ 58 & 0.33 &$\nodata$  &$\nodata$  &{\rm RR?}  & {\rm (c)}\\
    4328 &~~621     & 00~ 48~ 39.0 &+42~ 14~ 22 & 1.68?&22.52  &21.18  &{\rm CC}  & \\
    4420 &$\nodata$ & 00~ 48~ 38.9 &+42~ 19~ 45 & 0.21 &$\nodata$  &$\nodata$ &{\rm RR?}  & \\
    4435 &$\nodata$ & 00~ 48~ 38.9 &+42~ 21~ 45 & 0.26 &$\nodata$  &$\nodata$ &{\rm RR?}  & {\rm (c)}\\
    4472 &~1880     & 00~ 48~ 38.4 &+42~ 14~ 19 & 0.63 &23.73  &22.92  &{\rm CC}  & \\
    4475 &$\nodata$ & 00~ 48~ 38.6 &+42~ 18~ 42 & 0.39 &$\nodata$  &$\nodata$  &{\rm RR?}  & \\
    4562 &~~390     & 00~ 48~ 38.2 &+42~ 15~ 45 & 9.40 &21.48(21.74)&20.62(20.43) &{\rm CC}  &  {\rm V2~(f)}\\
    4564 &$\nodata$ & 00~ 48~ 38.3 &+42~ 17~ 27 & 0.62 &$\nodata$  &$\nodata$  &{\rm RRab?}  & \\
    4578 &$\nodata$ & 00~ 48~ 38.3 &+42~ 20~ 35 & 0.33 &$\nodata$  &$\nodata$  &{\rm CC}  & {\rm (c)}\\
    4592 &$\nodata$ & 00~ 48~ 38.3 &+42~ 19~ 54 & 0.26 &$\nodata$  &$\nodata$  &{\rm RR?}  & \\
    4705 &$\nodata$ & 00~ 48~ 38.0 &+42~ 21~ 57 & 2.62 &$\nodata$  &$\nodata$  &{\rm ?}  & {\rm (c)}\\
    4809 &$\nodata$ & 00~ 48~ 37.3 &+42~ 15~ 17 & 0.38 &$\nodata$  &$\nodata$  &{\rm RR?}  & \\
    4875 &$\nodata$ & 00~ 48~ 37.6 &+42~ 14~ 21 & 0.57 &$\nodata$  &$\nodata$  &{\rm ?}  & {\rm (d)}\\
    4936 &16336     & 00~ 48~ 36.9 &+42~ 15~ 25 & 0.57 &24.96  &24.31  &{\rm AC/spCC/RR}  &{\rm (o)} \\
    5033 &$\nodata$ & 00~ 48~ 36.7 &+42~ 17~ 10 & 0.21 &$\nodata$  &$\nodata$  &{\rm RR?}  & \\
    5089 &54011     & 00~ 48~ 36.4 &+42~ 14~ 57 & 0.605&25.75(25.65)&25.36(25.15)&{\rm RRab} & {\rm V1~(f)}\\
            \hline
            \end{array}
	    $$
\par\noindent
\end{table*}

  \begin{table*}
 \footnotesize
\tablenum{9}
      \caption[]{continued}
     $$
         \begin{array}{ccccccclc}
            \hline
           \noalign{\smallskip}

    5093 &~5960     & 00~ 48~ 36.7 &+42~ 20~ 05 & 0.36 &24.52  &24.57  &{\rm RR?/Bin.?/Be?}  & \\
    5169 &~~295     & 00~ 48~ 36.3 &+42~ 18~ 51 & 1.68?&21.33  &19.76  &{\rm CC}& \\
    5202 &~~442     & 00~ 48~ 36.2 &+42~ 19~ 17 & 0.51 &21.95  &20.28  &{\rm CC}& \\
    5205 &~~741     & 00~ 48~ 36.0 &+42~ 16~ 23 & 0.33 &22.92  &23.05  &{\rm CC?/Bin.?/Be?}& \\
    5305 &~~768     & 00~ 48~ 35.6 &+42~ 16~ 02 & 0.34 &22.97  &22.96  &{\rm CC?/Bin.?/Be?}& \\
    5384 &10225     & 00~ 48~ 35.4 &+42~ 19~ 13 & 0.33 &24.78  &24.84  &{\rm RR}  & \\
    5385 &27107     & 00~ 48~ 35.2 &+42~ 15~ 51 & 0.64 &25.22  &24.71  &{\rm RRab}  & \\
    5510 &$\nodata$ & 00~ 48~ 35.0 &+42~ 18~ 40 & 0.41 &$\nodata$  &$\nodata$  &{\rm RR} & {\rm (d)}\\
    5512 &$\nodata$ & 00~ 48~ 35.0 &+42~ 20~ 52 & 0.81 &$\nodata$  &$\nodata$  &{\rm AC/spCC/RR} & {\rm (c,o)}\\
    5847 &$\nodata$ & 00~ 48~ 33.6 &+42~ 15~ 07 & 0.45 &$\nodata$  &$\nodata$  &{\rm ?}  & {\rm (g)}\\
    5850 &24113     & 00~ 48~ 33.7 &+42~ 16~ 32 & 0.57 &25.14  &24.29  &{\rm RRab} & \\
    6191 &24896     & 00~ 48~ 32.7 &+42~ 19~ 41 & 1.15 &25.21  &25.24  &{\rm RRab} & \\
    6254 &$\nodata$ & 00~ 48~ 32.2 &+42~ 14~ 38 & 0.57 &$\nodata$  &$\nodata$  &{\rm RRab} & \\
    6269 &~3009     & 00~ 48~ 32.1 &+42~ 15~ 17 & 0.53 &24.09  &23.49  &{\rm AC/spCC/RR}  &{\rm (o)} \\
    6276 &$\nodata$ & 00~ 48~ 32.3 &+42~ 19~ 15 & 0.26 &$\nodata$  &$\nodata$  &{\rm RRc}  & \\
    6305 &$\nodata$ & 00~ 48~ 32.2 &+42~ 19~ 44 & 0.65 &$\nodata$  &$\nodata$  &{\rm CC}  & {\rm (e)}\\
    6375 &~~427     & 00~ 48~ 31.8 &+42~ 16~ 32 & 5.1: &22.13(21.98)  &21.47(21.57)  &{\rm CC}  & {\rm V4~(f)}\\
    6452 &$\nodata$ & 00~ 48~ 31.7 &+42~ 21~ 35 & 0.28 &$\nodata$  &$\nodata$  &{\rm RR?}  & {\rm (c)}\\
    6849 &~~430     & 00~ 48~ 30.3 &+42~ 16~ 46 & 0.64 &21.92  &20.53  &{\rm CC}  & \\
    6957 &$\nodata$ & 00~ 48~ 30.1 &+42~ 18~ 04 & 0.49 &$\nodata$  &$\nodata$  &{\rm RRab}  & \\
    7292 &~6174     & 00~ 48~ 28.4 &+42~ 18~ 18 & 0.77 &24.51	   &24.15	 &{\rm AC/spCC/RR}  &{\rm (o)} \\
    7378 &$\nodata$ & 00~ 48~ 28.1 &+42~ 20~ 49 & 1.12 &$\nodata$  &$\nodata$  &{\rm Bin.}  & {\rm (c)}\\
    7412 &$\nodata$ & 00~ 48~ 27.9 &+42~ 20~ 38 & 0.67 &$\nodata$  &$\nodata$  &{\rm CC?}  & {\rm (c,d)}\\
    7799 &~~245     & 00~ 48~ 26.4 &+42~ 18~ 28 & 0.41 &20.97  &19.39  &{\rm CC}  & {\rm (h)}\\
    7860 &~9033     & 00~ 48~ 26.1 &+42~ 17~ 19 & 1.82 &24.64  &23.40  &{\rm AC?/spCC?}  &{\rm (h,o)} \\
    7889 &$\nodata$ & 00~ 48~ 26.1 &+42~ 20~ 44 & 2.75 &$\nodata$  &$\nodata$  &{\rm CC}  & {\rm (c)}\\
    7897 &~~513     & 00~ 48~ 25.9 &+42~ 17~ 27 & 0.27 &22.23  &20.60  &{\rm ?}  & \\
    7931 &$\nodata$ & 00~ 48~ 25.8 &+42~ 21~ 34 & 0.60 &$\nodata$  &$\nodata$  &{\rm RRab}  & {\rm (c)}\\
    7961 &~1452     & 00~ 48~ 25.4 &+42~ 14~ 08 & 0.23 &23.58  &23.70  &{\rm AC?/ \beta ~Ceph.?/Bin.?}  & {\rm (o)}\\
    8097 &$\nodata$ & 00~ 48~ 24.8 &+42~ 14~ 30 & 0.40 &$\nodata$  &$\nodata$  &{\rm RR}  & \\
          \hline
            \end{array}
	    $$
\par\noindent
\end{table*}

  \begin{table*}
 \footnotesize
\tablenum{9}
      \caption[]{continued}
     $$
         \begin{array}{ccccccclc}
            \hline
           \noalign{\smallskip}

    8098 &$\nodata$ & 00~ 48~ 25.1 &+42~ 19~ 13 & 0.58 &$\nodata$  &$\nodata$  &{\rm ?}  & \\
    8106 &$\nodata$ & 00~ 48~ 24.8 &+42~ 14~ 28 & 0.48 &$\nodata$  &$\nodata$  &{\rm RRab}  & \\
    8318 &~1373     & 00~ 48~ 24.2 &+42~ 14~ 47 & 0.91 &23.53  &23.66  &{\rm AC?/Be?/Bin.?}  & {\rm (o)}\\
    8337 &$\nodata$ & 00~ 48~ 24.4 &+42~ 20~ 13 & 0.48 &$\nodata$  &$\nodata$  &{\rm AC?/RR?}  & {\rm (c,o)}\\
    8630 &$\nodata$ & 00~ 48~ 23.0 &+42~ 20~ 53 & 0.51 &$\nodata$  &$\nodata$  &{\rm CC?}  & {\rm (c)}\\
    8877 &~~986     & 00~ 48~ 21.9 &+42~ 17~ 38 & 0.56 &23.21  &23.21  &{\rm Bin.}  & \\
    8948 &~~933     & 00~ 48~ 21.6 &+42~ 16~ 55 & 3.25 &22.74(23.08)  &22.03(22.07)  &{\rm CC}  & {\rm V3~(f)}\\
    9070 &$\nodata$ & 00~ 48~ 21.2 &+42~ 18~ 39 & 4.35 &$\nodata$  &$\nodata$  &{\rm ?}  & {\rm (g)}\\
    9171 &~3836     & 00~ 48~ 21.0 &+42~ 19~ 39 & 0.574 &23.52(24.21)  &23.36(23.12)  &{\rm Bin.}  & {\rm V6~(f)}\\
    9231 &~~181     & 00~ 48~ 20.5 &+42~ 16~ 26 & 1.56? &20.56  &20.65  &{\rm CC?}  & \\
    9272 &11037     & 00~ 48~ 20.3 &+42~ 15~ 35 & 0.63 &24.83  &25.06  &{\rm RR?}  &{\rm (l)}\\
    9319 &$\nodata$ & 00~ 48~ 20.1 &+42~ 15~ 20 & 1.54 &$\nodata$  &$\nodata$  &{\rm CC}  & {\rm (e)}\\
    9413 &$\nodata$ & 00~ 48~ 20.1 &+42~ 21~ 36 & 0.21 &$\nodata$  &$\nodata$  &{\rm RRc?}  & {\rm (c)}\\
    9418 &$\nodata$ & 00~ 48~ 20.0 &+42~ 20~ 32 & 0.57 &$\nodata$  &$\nodata$  &{\rm RRab}  & {\rm (c)}\\
    9555 &$\nodata$ & 00~ 48~ 19.1 &+42~ 15~ 28 & 0.50 &$\nodata$  &$\nodata$  &{\rm CC?}  & {\rm (e)}\\
    9626 &$\nodata$ & 00~ 48~ 19.1 &+42~ 20~ 42 & 1.49 &$\nodata$  &$\nodata$  &{\rm AC?/spCC?}  & {\rm (c,o)}\\
    9717 &~~175     & 00~ 48~ 18.9 &+42~ 19~ 42 & 0.83 &20.41  &19.56  &{\rm CC}  & \\
    9780 &17186     & 00~ 48~ 18.3 &+42~ 15~ 41 & 0.58 &24.99  &24.53  &{\rm RR}  & \\
    9910 &~4972     & 00~ 48~ 17.9 &+42~ 17~ 37 & 0.24 &24.33  &23.03  &{\rm CC?}  & \\
   10134 &$\nodata$ & 00~ 48~ 17.2 &+42~ 21~ 41 & 2.64 &$\nodata$  &$\nodata$  &{\rm ?}  & {\rm (c)}\\ 
   10163 &$\nodata$ & 00~ 48~ 16.9 &+42~ 16~ 43 & 0.48 &$\nodata$  &$\nodata$  &{\rm RRab?}  & \\ 
   10287 &$\nodata$ & 00~ 48~ 16.4 &+42~ 18~ 47 & 0.44 &$\nodata$  &$\nodata$  &{\rm Bin.}  & \\
   10378 &~1310     & 00~ 48~ 15.9 &+42~ 19~ 37 & 1.59 &23.42  &22.43  &{\rm CC}  & \\ 
   10396 &$\nodata$ & 00~ 48~ 15.8 &+42~ 17~ 52 & 0.34 &$\nodata$  &$\nodata$  &{\rm RR?}  &{\rm (d)} \\ 
   10639 &~~387     & 00~ 48~ 14.8 &+42~ 19~ 35 & 0.41 &21.83  &22.00  &{\rm CC?/Be?}  & \\ 
   10835 &$\nodata$ & 00~ 48~ 13.8 &+42~ 18~ 55 & 0.54 &$\nodata$  &$\nodata$  &{\rm RRab}  & \\ 
   10988 &~~437     & 00~ 48~ 13.3 &+42~ 19~ 46 & 0.51 &21.94  &20.23  &{\rm CC}  & \\
   11028 &~~710     & 00~ 48~ 13.1 &+42~ 18~ 46 & 1.54 &22.71  &21.01  &{\rm CC}  & \\ 
   11733 &$\nodata$ & 00~ 48~ 10.8 &+42~ 20~ 09 & 0.63 &$\nodata$  &$\nodata$  &{\rm AC/spCC}  & {\rm (c,o)}\\ 
   11774 &~~424     & 00~ 48~ 10.6 &+42~ 19~ 13 & 1.89 &21.99  &22.00  &{\rm CC}  & \\ 
          \hline
            \end{array}
	    $$
\par\noindent
\end{table*}

  \begin{table*}
 \footnotesize
\tablenum{9}
      \caption[]{continued}
     $$
         \begin{array}{ccccccclc}
            \hline
           \noalign{\smallskip}

   $\nodata$ &~~480 & 00~ 48~ 10.2 &+42~ 16~ 31 & 2.92 &22.58(22.22)  &21.97(22.43)  &{\rm CC}  & {\rm V5~(f,i)}\\
   $\nodata$ &14532 & 00~ 48~ 43.4 &+42~ 08~ 23 &$\nodata$ &24.93 &24.80  &{\rm RR}  & {\rm (i)}\\
   $\nodata$ &10800 & 00~ 48~ 09.2 &+42~ 15~ 45 &$\nodata$ &24.82 &25.09  &{\rm RR}  & {\rm (i)}\\
   $\nodata$ &~5707 & 00~ 48~ 28.4 &+42~ 06~ 21 &$\nodata$ &24.46 &24.01  &{\rm AC/spCC/RR}  & {\rm (i,o)}\\
   $\nodata$ &~~502 & 00~ 48~ 23.2 &+42~ 13~ 29 &$\nodata$ &22.27 &21.78  &{\rm CC}  & {\rm (i)}\\
   $\nodata$ &~~146 & 00~ 48~ 35.5 &+42~ 13~ 37 &$\nodata$ &20.20 &20.19  &{\rm CC/Bin.}  & {\rm (i)}\\
          \hline
            \end{array}
	    $$
\par\noindent
\end{table*}

  \begin{table*}
  \footnotesize
      \tablenum{9}
      \caption[]{continued}
     $$
        \begin{array}{ccccccclc}
              \hline
              \hline
           \noalign{\smallskip}
           \multicolumn{9}{c}{\rm CCD2~-~Field~S2}      \\
           {\rm Id}  & {\rm Id}      &{\rm \alpha } & {\rm \delta} & {\rm P}      & $B$         & $V$        & {\rm Type}& {\rm Notes}\\
           {\rm (ISIS)}& {\rm (DoPHOT)}&{\rm (2000)}  & {\rm (2000)} & {\rm (days)} & {\rm (mag)} & {\rm (mag)} & &\\
                           &       {\rm (a)}   &              &              &              & {\rm (b)}   & {\rm (b)}   &          &\\
          \noalign{\smallskip}
            \hline
          \noalign{\smallskip}
     5532 &10706    & 00~ 49~ 27.7 &+42~ 06~ 18 & 0.18 &25.24	  &25.18 &{\rm RRc?} &\\ 
     5533 &~~~80    & 00~ 49~ 28.7 &+42~ 19~ 47 & 0.40 &19.62	  &18.90 &{\rm Bin.} &\\ 
     5542 &$\nodata$& 00~ 49~ 28.6 &+42~ 20~ 43 & 0.28 &$\nodata$ &$\nodata$ &{\rm RRc} &{\rm (c)}\\ 
     5612 &$\nodata$& 00~ 49~ 27.8 &+42~ 21~ 45 & 0.38 &$\nodata$ &$\nodata$ &{\rm AC/spCC}&{\rm (c,o)}\\ 
     5687 &~~379    & 00~ 49~ 26.2 &+42~ 06~ 21 & 0.60 &22.41	  &20.88	&{\rm CC} &\\ 
     5721 &$\nodata$& 00~ 49~ 26.9 &+42~ 19~ 55 & 2.93 &$\nodata$ &$\nodata$ &{\rm ?} &\\ 
     5743 &~~215    & 00~ 49~ 26.5 &+42~ 15~ 31 & 1.81 &21.23	  &19.94	  &{\rm CC} &\\ 
     5796 &~~190    & 00~ 49~ 25.8 &+42~ 09~ 55 & 0.34 &21.04	  &20.33       &{\rm CC} &\\ 
     5838 &11247    & 00~ 49~ 25.7 &+42~ 13~ 02 & 0.40 &25.23	  &24.64     &{\rm RRc} &\\ 
     5845 &$\nodata$& 00~ 49~ 26.1 &+42~ 19~ 42 & 0.44 &$\nodata$ &$\nodata$ &{\rm RR} &\\ 
     5864 &$\nodata$& 00~ 49~ 26.1 &+42~ 20~ 29 & 0.39 &$\nodata$ &$\nodata$ &{\rm RRc}&{\rm (c)}\\ 
     5992 &~~738    & 00~ 49~ 24.3 &+42~ 05~ 33 & 0.33 &23.47	  &22.77	  &{\rm AC?/Bin.?} &{\rm (o)}\\ 
     6062 &$\nodata$& 00~ 49~ 24.0 &+42~ 06~ 10 & 0.16 &$\nodata$ &$\nodata$ &{\rm RR?/Bin.?} &\\ 
     6093 &~~281    & 00~ 49~ 24.1 &+42~ 10~ 21 & 0.40 &21.86	  &21.27	  &{\rm ?} &\\ 
     6135 &$\nodata$& 00~ 49~ 24.6 &+42~ 21~ 59 & 0.32 &$\nodata$ &$\nodata$ &{\rm ?} &{\rm (c)}\\ 
     6166 &$\nodata$& 00~ 49~ 24.4 &+42~ 21~ 47 & 0.12 &$\nodata$ &$\nodata$ &{\rm RR} &{\rm (c)}\\ 
     6175 &12057    & 00~ 49~ 24.1 &+42~ 18~ 42 & 0.47 &25.25	  &24.52	&{\rm RRab} &\\ 
     6201 &$\nodata$& 00~ 49~ 24.2 &+42~ 21~ 33 & 0.41 &$\nodata$ &$\nodata$ &{\rm RR} &{\rm (c)}\\ 
     6250 &$\nodata$& 00~ 49~ 23.9 &+42~ 21~ 54 & 0.20 &$\nodata$ &$\nodata$ &{\rm RRc}&{\rm (c)}\\ 
     6266 &~~232    & 00~ 49~ 23.3 &+42~ 14~ 08 & 0.25 &21.48	  &20.78	  &{\rm CC} &\\ 
     6289 &~~243    & 00~ 49~ 23.3 &+42~ 17~ 04 & 0.65 &21.53	  &20.16	&{\rm CC} &\\ 
     6313 &~~255    & 00~ 49~ 22.6 &+42~ 07~ 18 & 0.16 &21.61	  &20.43     &{\rm CC?} &\\ 
     6371 &~~188    & 00~ 49~ 22.5 &+42~ 10~ 18 & 0.78 &21.07	  &20.82     &{\rm ?} &\\ 
     6394 &~~210    & 00~ 49~ 22.9 &+42~ 16~ 21 & 0.63 &21.22	  &19.89       &{\rm CC?} &\\ 
     6510 &13187    & 00~ 49~ 22.2 &+42~ 15~ 24 & 0.31 &25.25	  &23.83     &{\rm Bin.?/RRc?} &\\ 
     6568 &12477    & 00~ 49~ 21.3 &+42~ 05~ 56 & 0.32 &25.30	  &24.95     &{\rm RRc} &\\ 
            \hline
            \end{array}
	    $$
\par\noindent
\end{table*}

  \begin{table*}
 \footnotesize
\tablenum{9}
      \caption[]{continued}
     $$
        \begin{array}{ccccccclc}
             \hline
           \noalign{\smallskip}

     6623 &~~217    & 00~ 49~ 21.3 &+42~ 10~ 09 & 0.65 &21.34	  &21.23     &{\rm CC} &\\ 
     6645 &5323     & 00~ 49~ 20.9 &+42~ 05~ 37 & 0.34 &24.90	  &24.61     &{\rm AC/spCC/RR} &{\rm (o)}\\ 
     6703 &18856    & 00~ 49~ 20.9 &+42~ 09~ 15 & 0.59 &25.45	  &24.36       &{\rm RRab} &\\ 
     6825 &~~223    & 00~ 49~ 20.6 &+42~ 12~ 42 & 4.86 &21.43	  &21.01	&{\rm CC} &\\ 
     6826 &$\nodata$& 00~ 49~ 20.7 &+42~ 14~ 15 & 0.32 &$\nodata$ &$\nodata$ &{\rm RRc} &\\ 
     6957 &$\nodata$& 00~ 49~ 19.8 &+42~ 08~ 03 & 0.17 &$\nodata$ &$\nodata$ &{\rm RR?} &\\ 
     6992 &10172    & 00~ 49~ 19.7 &+42~ 08~ 33 & 0.33 &25.26	  &25.82	&{\rm RRab} &\\ 
     7009 &~3573    & 00~ 49~ 20.3 &+42~ 19~ 35 & 0.36? &24.79     &25.45     &{\rm RRab} &\\ 
     7017 &$\nodata$& 00~ 49~ 20.6 &+42~ 11~ 20 & 0.33 &$\nodata$ &$\nodata$ &{\rm Bin.} &\\ 
     7077 &$\nodata$& 00~ 49~ 20.2 &+42~ 22~ 24 & 0.25 &$\nodata$ &$\nodata$ &{\rm Bin.} &{\rm (c)}\\ 
     7189 &~7510    & 00~ 49~ 18.7 &+42~ 06~ 30 & 0.76 &25.09	  &25.15     &{\rm RRab} &\\ 
     7241 &~~150    & 00~ 49~ 18.8 &+42~ 13~ 08 & 0.66 &20.66	  &19.19     &{\rm Bin.?/CC?} &\\ 
     7403 &~~285    &  00~ 49~ 18.1 &+42~ 17~ 11 & 0.34 &21.81    &20.05	 &{\rm CC} &\\ 
     7437 &~~284    &  00~ 49~ 17.7 &+42~ 12~ 28 & 0.39 &21.82    &20.32	 &{\rm ?} &\\ 
     7455 &$\nodata$&  00~ 49~ 18.1 &+42~ 19~ 57 & 0.26 &$\nodata$&$\nodata$&{\rm RRc} &\\ 
     7470 &~2143    &  00~ 49~ 17.4 &+42~ 09~ 10 & 0.79 &24.46    &24.40       &{\rm AC?/spCC?/RR?} &{\rm (o)}\\ 
     7476 &~9568    &  00~ 49~ 17.7 &+42~ 15~ 09 & 0.33 &25.15    &24.55       &{\rm RR} &\\ 
     7498 &$\nodata$&  00~ 49~ 17.1 &+42~ 05~ 34 & 0.34 &$\nodata$&$\nodata$&{\rm RRc}&\\ 
     7524 &$\nodata$&  00~ 49~ 18.0 &+42~ 21~ 27 & 0.39 &$\nodata$&$\nodata$&{\rm RRc}&{\rm (c)}\\ 
     7593 &$\nodata$&  00~ 49~ 16.9 &+42~ 10~ 46 & 0.57 &$\nodata$&$\nodata$&{\rm RRab}&\\ 
     7594 &$\nodata$&  00~ 49~ 16.9 &+42~ 11~ 44 & 0.40 &$\nodata$&$\nodata$&{\rm RR?} &\\ 
     7622 &~6055    &  00~ 49~ 16.5 &+42~ 06~ 08 & 0.16 &24.96    &24.53      &{\rm RRc} &\\ 
     7628 &~~227    &  00~ 49~ 16.7 &+42~ 11~ 15 & 0.66 &21.39    &19.90       &{\rm CC?} &\\ 
     7651 &~~231    &  00~ 49~ 16.8 &+42~ 17~ 48 & 0.33 &21.43    &19.91       &{\rm CC?} &\\ 
     7662 &~~132    &  00~ 49~ 16.4 &+42~ 10~ 59 & 0.25 &20.53    &20.10	 &{\rm CC} &\\ 
     7666 &$\nodata$&  00~ 49~ 16.5 &+42~ 14~ 09 & 0.25?&$\nodata$&$\nodata$&{\rm CC} &{\rm (e)}\\ 
     7688 &~6354    &  00~ 49~ 16.1 &+42~ 05~ 58 & 0.33 &25.00    &24.90	 &{\rm RRc?/Bin.?} &\\ 
     7697 &~~136    &  00~ 49~ 16.7 &+42~ 18~ 02 & 0.29 &20.56    &20.00       &{\rm ?} &\\ 
     7771 &$\nodata$&  00~ 49~ 16.6 &+42~ 22~ 18 & 0.16 &$\nodata$&$\nodata$&{\rm Bin.} &{\rm (c)}\\ 
     7793 &$\nodata$&  00~ 49~ 16.2 &+42~ 20~ 00 & 0.38 &$\nodata$&$\nodata$&{\rm Bin.} &\\ 
            \hline
            \end{array}
	    $$
\par\noindent
\end{table*}

  \begin{table*}
  \footnotesize
\tablenum{9}
      \caption[]{continued}
     $$
         \begin{array}{rcccccclc}
           \hline
           \noalign{\smallskip}

     7852 &~8382    &  00~ 49~ 15.8 &+42~ 15~ 00 & 0.35 &25.13    &25.10	 &{\rm RR} &\\
     7869 &~4129    &  00~ 49~ 15.7 &+42~ 15~ 08 & 0.45 &24.83    &24.90	 &{\rm RR} &\\
     7913 &~~184    &  00~ 49~ 15.3 &+42~ 14~ 12 & 0.25?&20.95    &19.97   &{\rm CC} &\\
     7989 &$\nodata$&  00~ 49~ 15.2 &+42~ 15~ 48 & 0.33 &$\nodata$&$\nodata$&{\rm RRc} &\\
     8003 &~~273    &  00~ 49~ 15.0 &+42~ 14~ 10 & 0.25 &21.77    &20.58       &{\rm CC} &\\
     8113 &$\nodata$&  00~ 49~ 14.9 &+42~ 21~ 08 & 0.26 &$\nodata$&$\nodata$&{\rm RRc} &{\rm (c)}\\
     8119 &~5400    &  00~ 49~ 14.2 &+42~ 09~ 29 & 0.48 &24.89    &24.28	 &{\rm AC/spCC/RR} &{\rm (o)}\\
     8155 &$\nodata$&  00~ 49~ 13.9 &+42~ 06~ 21 & 0.33 &$\nodata$&$\nodata$&{\rm RR?} &\\
     8315 &$\nodata$&  00~ 49~ 13.2 &+42~ 05~ 40 & 0.40 &$\nodata$&$\nodata$&{\rm RR?} &\\
     8495 &~~139    &  00~ 49~ 13.4 &+42~ 15~ 45 & 0.25 &20.55    &19.45	 &{\rm CC?} &\\
     8558 &$\nodata$&  00~ 49~ 13.0 &+42~ 14~ 48 & 0.28? &$\nodata$&$\nodata$&{\rm CC} &{\rm (m)}\\
     8733 &$\nodata$&  00~ 49~ 12.9 &+42~ 21~ 30 & 0.44 &$\nodata$&$\nodata$&{\rm Bin.} &{\rm (c)}\\
     8961 &$\nodata$&  00~ 49~ 12.3 &+42~ 20~ 44 & 0.31 &$\nodata$&$\nodata$&{\rm RR} &{\rm (c)}\\
     9136 &$\nodata$&  00~ 49~ 11.5 &+42~ 17~ 21 & 0.64 &$\nodata$&$\nodata$&{\rm CC?} &{\rm (e)}\\
     9240 &~~248    &  00~ 49~ 11.1 &+42~ 14~ 17 & 0.25 &21.64    &21.50	 &{\rm CC} &\\
     9311 &12765    &  00~ 49~ 10.9 &+42~ 16~ 02 & 0.36 &25.33    &25.22       &{\rm RR} &\\
     9332 &$\nodata$&  00~ 49~ 10.9 &+42~ 17~ 57 & 0.34 &$\nodata$&$\nodata$&{\rm RR} &\\
     9365 &$\nodata$ &  00~ 49~ 10.5 &+42~ 12~ 28 & 0.20 &$\nodata$ &$\nodata$ &{\rm RRc} &\\
     9413 &$\nodata$ &  00~ 49~ 10.2 &+42~ 11~ 25 & 0.39 &$\nodata$ &$\nodata$ &{\rm RRc} &\\
     9418 &$\nodata$ &  00~ 49~ 10.0 &+42~ 09~ 12 & 0.34 &$\nodata$ &$\nodata$ &{\rm RR} &\\
     9604 &$\nodata$ &  00~ 49~ 09.3 &+42~ 14~ 26 & 0.25?&$\nodata$ &$\nodata$ &{\rm CC} &{\rm (e)}\\
     9608 &$\nodata$ &  00~ 49~ 09.0 &+42~ 08~ 33 & 0.63 &$\nodata$ &$\nodata$ &{\rm RRab} &\\
     9620 &$\nodata$ &  00~ 49~ 09.3 &+42~ 16~ 14 & 0.56 &$\nodata$ &$\nodata$ &{\rm RRab} &\\
     9646 &~~113     &  00~ 49~ 09.1 &+42~ 12~ 56.& 0.63 &20.16 	 &19.26     &{\rm CC} &\\
     9650 &$\nodata$ &  00~ 49~ 09.3 &+42~ 17~ 16 & 0.16 &$\nodata$ &$\nodata$ &{\rm RR} &\\
     9652 &$\nodata$ &  00~ 49~ 09.4 &+42~ 20~ 01 & 0.33 &$\nodata$ &$\nodata$ &{\rm RR} &\\
     9675 & ~~127    &  00~ 49~ 08.9 &+42~ 14~ 31 & 0.20?&20.41 	 &19.18     &{\rm CC} &\\
     9687 &$\nodata$ &  00~ 49~ 08.4 &+42~ 05~ 33 & 0.21 &$\nodata$ &$\nodata$ &{\rm RR} &\\
     9702 &~~219     &  00~ 49~ 08.6 &+42~ 10~ 14 & 0.20?&21.36       &20.61	 &{\rm CC} &\\
     9723 &$\nodata$ &  00~ 49~ 08.4 &+42~ 05~ 59 & 0.26 &$\nodata$ &$\nodata$ &{\rm RR?} &\\
            \hline
            \end{array}
	    $$
\par\noindent
\end{table*}

  \begin{table*}
  \footnotesize
\tablenum{9}
      \caption[]{continued}
     $$
         \begin{array}{ccccccclc}
           \hline
           \noalign{\smallskip}

     9792 &$\nodata$ &  00~ 49~ 08.5 &+42~ 16~ 17 & 0.33 &$\nodata$ &$\nodata$ &{\rm RRc} &\\
     9853 &$\nodata$ &  00~ 49~ 08.0 &+42~ 14~ 35 & 0.78 &$\nodata$ &$\nodata$ &{\rm RR?} &\\
     9945 &12894     &  00~ 49~ 07.4 &+42~ 10~ 50 & 0.46 &25.35       &25.51	 &{\rm RRab} &\\
     9968 &~5586     &  00~ 49~ 07.3 &+42~ 10~ 57 & 0.43 &24.97 	 &25.26     &{\rm RR?} &\\
     9988 &~~172     &  00~ 49~ 07.3 &+42~ 12~ 35 & 0.34?&20.85       &19.47	 &{\rm CC?} &\\
    10104 &10268     &  00~ 49~ 06.6 &+42~ 07~ 47 & 0.66 &25.19       &24.65	 &{\rm RRab} &\\
    10112 &$\nodata$ &  00~ 49~ 06.6 &+42~ 07~ 30 & 0.27 &$\nodata$ &$\nodata$ &{\rm RRc} &\\
    10125 &~~169     &  00~ 49~ 07.0 &+42~ 15~ 56 & 0.34?&20.87 	 &20.08     &{\rm CC} &\\
    10161 &$\nodata$ &  00~ 49~ 06.9 &+42~ 19~ 07 & 0.21 &$\nodata$ &$\nodata$ &{\rm RRc} &\\
    10244 &~7838     &  00~ 49~ 06.1 &+42~ 10~ 13 & 0.34 &25.04       &24.14	 &{\rm ?} &\\
    10261 &$\nodata$ &  00~ 49~ 06.2 &+42~ 14~ 09 & 0.33?&$\nodata$ &$\nodata$ &{\rm CC} &{\rm(e)}\\
    10283 &$\nodata$ &  00~ 49~ 06.5 &+42~ 21~ 32 & 0.36?&$\nodata$ &$\nodata$ &{\rm RRab} &{\rm (c)}\\
    10300 &$\nodata$ &  00~ 49~ 05.7 &+42~ 06~ 43 & 0.19 &$\nodata$ &$\nodata$ &{\rm RR} &\\
    10334 &$\nodata$ &  00~ 49~ 06.0 &+42~ 14~ 35 & 0.16 &$\nodata$ &$\nodata$ &{\rm RRc} &\\
    10357 &~~205     &  00~ 49~ 05.8 &+42~ 12~ 21 & 0.25?&21.14       &20.00	 &{\rm ?} &\\
    10365 &~~189     &  00~ 49~ 05.8 &+42~ 15~ 31 & 0.25?&21.01     &20.00     &{\rm CC} &\\
    10490 &10405     &  00~ 49~ 04.8 &+42~ 06~ 35 & 0.38 &25.24        &25.35	  &{\rm RRc} &\\
    10494 &~~209     &  00~ 49~ 05.4 &+42~ 18~ 21 & 0.64 &21.27        &20.73	  &{\rm CC?} &\\
    10506 &$\nodata$ &  00~ 49~ 05.6 &+42~ 21~ 44 & 0.45 &$\nodata$ &$\nodata$ &{\rm RRab} &{\rm (c)}\\
    10515 &$\nodata$ &  00~ 49~ 04.8 &+42~ 09~ 30 & 0.53 &$\nodata$ &$\nodata$ &{\rm RRab} &\\
    10525 &$\nodata$ &  00~ 49~ 04.7 &+42~ 07~ 22 & 0.20 &$\nodata$ &$\nodata$ &{\rm RR} &\\
    10536 &$\nodata$ &  00~ 49~ 05.3 &+42~ 21~ 23 & 1.76 &$\nodata$ &$\nodata$ &{\rm ?} &\\
    10577 &$\nodata$ & 00~ 49~ 04.6 &+42~ 12~ 31 & 0.64 &$\nodata$ &$\nodata$ &{\rm RRab} &\\
    10626 &$\nodata$ & 00~ 49~ 04.0 &+42~ 05~ 48 & 0.33 &$\nodata$ &$\nodata$ &{\rm RR} &\\
    10648 &~~119     & 00~ 49~ 04.2 &+42~ 14~ 02 & 0.25?&20.36        &19.85	   &{\rm CC?} &\\
    10653 &$\nodata$ & 00~ 49~ 04.6 &+42~ 21~ 01 & 0.33 &$\nodata$ &$\nodata$ &{\rm AC?/spCC?/RR?} &{\rm (c,o)}\\
    10666 &~1162     & 00~ 49~ 04.0 &+42~ 12~ 22 & 0.12 &23.90        &22.36	   &{\rm ?} &\\
    10680 &$\nodata$ & 00~ 49~ 03.7 &+42~ 05~ 44 & 0.22 &$\nodata$ &$\nodata$ &{\rm Bin.} &{\rm (e)}\\
    10688 &~3205     & 00~ 49~ 03.6 &+42~ 05~ 39 & 0.38 &24.58        &22.97	   &{\rm AC?/spCC?/RR?} &{\rm (o)}\\
    10756 &~~214     & 00~ 49~ 03.6 &+42~ 14~ 19 & 0.25?&21.27        &19.98	 &{\rm CC} &\\
            \hline
            \end{array}
	    $$
\par\noindent
\end{table*}

  \begin{table*}
  \footnotesize
\tablenum{9}
      \caption[]{continued}
     $$
         \begin{array}{ccccccclc}
           \hline
           \noalign{\smallskip}

    10780 &~~218     & 00~ 49~ 03.5 &+42~ 14~ 22 & 0.34?&21.31        &20.14	 &{\rm CC?} &\\
    10791 &~7226     & 00~ 49~ 03.5 &+42~ 14~ 15 & 0.67 &25.05        &24.83	 &{\rm RRab} &\\
    10992 &~~194     & 00~ 49~ 02.6 &+42~ 19~ 52 & 0.34?&21.02     &19.54	&{\rm CC} &\\
    10995 &$\nodata$ & 00~ 49~ 02.5 &+42~ 17~ 45 & 0.32 &$\nodata$ &$\nodata$ &{\rm RR} &\\
    11082 &~6532     & 00~ 49~ 01.7 &+42~ 19~ 08 & 4.31?&25.04        &25.44	 &{\rm ?} &\\
    11106 &~1495     & 00~ 49~ 01.5 &+42~ 17~ 13 & 0.86 &24.12        &22.75	 &{\rm CC} &\\
    11211 &$\nodata$ & 00~ 49~ 00.0 &+42~ 07~ 31 & 0.69 &$\nodata$ &$\nodata$ &{\rm RR} &\\
    11282 &$\nodata$ & 00~ 48~ 59.7 &+42~ 15~ 08 & 0.15 &$\nodata$ &$\nodata$ &{\rm Bin.?/RR?} &\\
    11291 &23265     & 00~ 48~ 59.8 &+42~ 17~ 44 & 0.38 &25.57     &24.52     &{\rm Bin.?} &\\
    11329 &$\nodata$ & 00~ 48~ 59.2 &+42~ 14~ 21 & 0.52 &$\nodata$ &$\nodata$ &{\rm RRab} &\\
    11383 &$\nodata$ & 00~ 48~ 58.6 &+42~ 13~ 08 & 0.51 &$\nodata$ &$\nodata$ &{\rm RRab} &\\
    11492 &~4457     & 00~ 48~ 57.7 &+42~ 15~ 37 & 0.66 &24.87        &25.15	   &{\rm RRab} &\\
    11497 &~2104     & 00~ 48~ 57.3 &+42~ 06~ 44 & 0.53 &24.41     &23.67     &{\rm AC/spCC/RR} &{\rm (o)}\\
    11588 &$\nodata$ & 00~ 48~ 56.6 &+42~ 06~ 02 & 0.38 &$\nodata$ &$\nodata$ &{\rm ?} &\\
    11681 &$\nodata$ & 00~ 48~ 56.0 &+42~ 10~ 05 & 2.85 &$\nodata$ &$\nodata$ &$\nodata$	&{\rm (n)}\\
    11763 &21067     & 00~ 48~ 55.4 &+42~ 14~ 12 & 0.46 &25.55        &25.01	   &{\rm RRab} &\\
    11812 &$\nodata$ & 00~ 48~ 55.3 &+42~ 19~ 06 & 0.27?&$\nodata$ &$\nodata$ &{\rm RRab} &\\
    11858 &$\nodata$ & 00~ 48~ 54.8 &+42~ 16~ 10 & 0.35 &$\nodata$ &$\nodata$ &{\rm RR} &{\rm (d)}\\
    11892 &~3094     & 00~ 48~ 54.5 &+42~ 13~ 23 & 0.65 &24.73     &25.53	&{\rm ?} &\\
    11935 &$\nodata$ & 00~ 48~ 53.9 &+42~ 06~ 07 & 0.25 &$\nodata$ &$\nodata$ &{\rm RR} &\\
    11955 &~2845     & 00~ 48~ 53.9 &+42~ 08~ 36 & 0.47 &24.64     &24.88	&{\rm RRab} &\\
    12000 &$\nodata$ & 00~ 48~ 53.5 &+42~ 05~ 54 & 0.82 &$\nodata$ &$\nodata$ &{\rm Bin.?/RR?} &\\
    12124 &~~138     & 00~ 48~ 52.8 &+42~ 08~ 40 & 0.12?&20.56        &19.84	 &{\rm ?} &\\
    12158 &$\nodata$ & 00~ 48~ 52.6 &+42~ 06~ 25 & 0.49 &$\nodata$ &$\nodata$ &{\rm RR?} &\\
    12214 &~~163     & 00~ 48~ 52.8 &+42~ 18~ 15 & 0.33 &20.84        &20.09	   &{\rm CC} &\\
    12291 &~7005     & 00~ 48~ 52.1 &+42~ 09~ 43 & 0.39 &25.03     &24.79	&{\rm RRc} &\\
    12408 &~3988     & 00~ 48~ 51.7 &+42~ 18~ 21 & 0.53 &24.83     &25.41	&{\rm RRab} &\\
            \hline
            \end{array}
	    $$
\par\noindent
\end{table*}

\clearpage

  \begin{table*}
  \footnotesize
\tablenum{9}
      \caption[]{continued}
     $$
         \begin{array}{rcccccclc}
           \hline
           \hline
           \noalign{\smallskip}
           \multicolumn{9}{c}{\rm CCD2~-~Field~H1}       \\
           {\rm Id~~} & {\rm Id}      &{\rm \alpha } & {\rm \delta} & {\rm P}      & $B$         & $V$        & {\rm Type}& {\rm Notes}\\
           {\rm (ISIS)}& {\rm (DoPHOT)}&{\rm (2000)}  & {\rm (2000)} & {\rm (days)} & {\rm (mag)} & {\rm (mag)} & &\\
                          &       {\rm (a)}   &              &              &              & {\rm (b)}   & {\rm (b)}   &          &\\
            \noalign{\smallskip}
            \hline
            \noalign{\smallskip}

    728  &$\nodata$ &00~ 48~ 32.5 &+40~ 20~ 11 & 0.32 &$\nodata$  &$\nodata$ &{\rm RR?} &\\ 
    730  &$\nodata$ &00~ 48~ 32.7 &+40~ 24~ 36 & 0.84 &$\nodata$  &$\nodata$ &{\rm ?} &{\rm (c)}\\
   1128  &$\nodata$ &00~ 48~ 31.3 &+40~ 20~ 06 & 0.32 &$\nodata$  &$\nodata$ &{\rm ?} &{\rm (n)}\\
   1212  &$\nodata$ &00~ 48~ 31.3 &+40~ 23~ 51 & 0.24 &$\nodata$  &$\nodata$ &{\rm Bin.} &{\rm (c)}\\
   1225  &$\nodata$ &00~ 48~ 31.0 &+40~ 20~ 00 & 0.25 &$\nodata$  &$\nodata$ &{\rm RR?} &\\
   1349  &~~546     &00~ 48~ 30.6 &+40~ 18~ 35 & 0.24 &22.06	  &22.05       &{\rm CC?} &\\
   1565  &$\nodata$ &00~ 48~ 29.8 &+40~ 18~ 08 & 0.44 &$\nodata$  &$\nodata$ &{\rm RR}  &\\ 
   1570  &$\nodata$ &00~ 48~ 29.9 &+40~ 20~ 49 & 0.25 &$\nodata$  &$\nodata$ &{\rm CC} & \\ 
   1578  &34061     &00~ 48~ 29.9 &+40~ 19~ 59 & 0.32 &26.52	  &24.81	  &{\rm RR} &\\ 
   1856  &$\nodata$ &00~ 48~ 29.2 &+40~ 23~ 51 & 0.44 &$\nodata$  &$\nodata$ &{\rm RR?} &{\rm (c)}\\ 
   1882  &~~341     &00~ 48~ 28.8 &+40~ 19~ 10 & 0.25 &20.85	  &20.48	  &{\rm CC} &\\
   2647  &$\nodata$ &00~ 48~ 26.1 &+40~ 23~ 21 & 0.24 &$\nodata$  &$\nodata$ &{\rm ?} &{\rm (c)}\\
   3046  &~~378     &00~ 48~ 23.8 &+40~ 19~ 51 & 0.32 &21.23	  &21.09	  &{\rm CC?/Be?} &\\
   3798  &$\nodata$ &00~ 48~ 19.6 &+40~ 24~ 25 & 0.34 &$\nodata$  &$\nodata$ &{\rm CC} &{\rm (c)}\\
   3821  &15025     &00~ 48~ 19.1 &+40~ 19~ 38 & 0.55 &25.84	  &24.63	  &{\rm RRab} &\\
   3938  &12921     &00~ 48~ 18.4 &+40~ 19~ 22 & 0.61 &25.73	  &24.73	  &{\rm RRab} &\\
   4054  &$\nodata$ &00~ 48~ 18.0 &+40~ 22~ 15 & 0.25 &$\nodata$  &$\nodata$ &{\rm RR} &\\
   4115  &~6455     &00~ 48~ 17.6 &+40~ 19~ 58 & 0.45 &25.17	  &24.32	  &{\rm RR} &\\
   4149  &~~438     &00~ 48~ 17.3 &+40~ 17~ 29 & 0.26 &21.70	  &20.08	  &{\rm CC?/Bin.?} &\\
   4547  &$\nodata$ &00~ 48~ 15.7 &+40~ 23~ 41 & 0.24 &$\nodata$  &$\nodata$ &{\rm Bin.} &{\rm (c)}\\
   4830  &$\nodata$ &00~ 48~ 14.4 &+40~ 23~ 00 & 0.32 &$\nodata$  &$\nodata$ &{\rm RR}   &{\rm (c)}\\
   5129  &$\nodata$ &00~ 48~ 13.1 &+40~ 21~ 26 & 0.28 &$\nodata$  &$\nodata$ &{\rm RR} &\\
   5302  &$\nodata$ &00~ 48~ 12.4 &+40~ 19~ 29 & 0.42 &$\nodata$  &$\nodata$ &{\rm RR?} &\\
   5585  &$\nodata$ &00~ 48~ 11.4 &+40~ 21~ 02 & 0.25 &$\nodata$  &$\nodata$ &{\rm RR?} &\\
   6007  &$\nodata$ &00~ 48~ 09.4 &+40~ 24~ 07 & 0.24 &$\nodata$  &$\nodata$ &{\rm RR} &{\rm (c)}\\
   6530  &20404     &00~ 48~ 06.6 &+40~ 20~ 37 & 0.56 &26.07	  &24.60	  &{\rm RR} &\\
           \hline
             \end{array}
	    $$
\par\noindent
\end{table*}

  \begin{table*}
  \footnotesize
\tablenum{9}
      \caption[]{continued}
     $$
         \begin{array}{ccccccclc}
           \hline
           \noalign{\smallskip}

   6881  &$\nodata$ &00~ 48~ 05.6 &+40~ 23~ 27 & 0.24 &$\nodata$  &$\nodata$ &{\rm Bin.} &{\rm (c)}\\
   7062  &~3629     &00~ 48~ 04.6 &+40~ 20~ 32 & 0.54 &24.77	  &25.22	&{\rm RRab} &\\
   7205  &~6453     &00~ 48~ 03.7 &+40~ 19~ 04 & 0.55 &25.21	  &25.09     &{\rm RRab} &\\
   7405  &$\nodata$ &00~ 48~ 02.7 &+40~ 21~ 14 & 0.40 &$\nodata$  &$\nodata$ &{\rm RR} &\\
   7533  &~~284     &00~ 48~ 01.8 &+40~ 18~ 01 & 0.20 &20.06	  &19.09	  &{\rm CC} &\\
   7781  &15531     &00~ 48~ 00.4 &+40~ 18~ 53 & 0.78 &25.93	  &25.70	&{\rm RRab} &\\
   8308  &$\nodata$ &00~ 47~ 58.0 &+40~ 23~ 35 & 0.34 &$\nodata$  &$\nodata$ &{\rm RR} &{\rm (c)}\\
           \hline
             \end{array}
	    $$
\par\noindent
\end{table*}

  \begin{table*}
 \footnotesize
      \caption[]{Identification and properties of confirmed variable stars in the M31 field S2}
         \label{lbt_var}
     $$
         \begin{array}{lccclllcclll}
	    \hline
            \hline
           \noalign{\smallskip}
           {\rm Name} & {\rm Id}  & {\rm \alpha } & {\rm \delta} & ~~ {\rm Type} &~~~~{\rm P}   & 
	    ~~~{\rm Epoch (a)}  & {\rm N_V} & {\rm N_B} & ~~{\rm \langle B\rangle}   
	    & ~{\rm A_B}& ~~{\rm \langle V\rangle}\\
            ~~        &  {\rm (b)}             &{\rm (2000)}   & {\rm (2000)} &             & ~{\rm (days)}&
	     ($-$2450000)        &             &             &~~{\rm (c)}&{\rm (mag)} &~~{\rm (d)}\\
            \noalign{\smallskip}
            \hline
            \noalign{\smallskip}
	    
{\rm ~V1} &~5089& 00~48~36.4&+42~14~57& {\rm RRab }	 &~0.605 &~4386.822   & 4 & 31  & 25.75& 1.03& 25.36\\  
{\rm ~V2} &~4562& 00~48~38.2&+42~15~45& {\rm Cepheid}	 &~9.40~ &~4385.200   & 5 & 53  & 21.48& 0.88& 20.62\\  
{\rm ~V3} &~8948& 00~48~21.6&+42~16~55& {\rm Cepheid}	 &~3.25~ &~4387.942   & 5 & 49  & 22.74& 1.07& 22.03\\  
{\rm ~V4} &~6375& 00~48~31.8&+42~16~32& {\rm Cepheid}	 &~5.1:~ &~4388.400:  & 5 & 51  & 22.13:&0.84:&21.47:\\ 
{\rm ~V5} &$\nodata$& 00~48~10.2&+42~16~31&{\rm Cepheid} &~2.92~ &~4383.700   & 5 & 51  & 22.58& 1.29& 21.97\\  
{\rm ~V6} &~9171& 00~48~21.0&+42~19~39& {\rm Binary}	 &~0.574 &~4389.790   & 3 & 49  & 23.52& 1.35& 23.36\\  
\hline
            \end{array}
	    $$
{\small $^{\mathrm{a}}$ Epochs correspond to the time of maximum light for the pulsating variables, to the time of the main minimum light
for the binary system.}\\
{\small $^{\mathrm{b}}$ Identification numbers in column 2 correspond to the ISIS IDs (see Table~\ref{candvar_coord}).}\\
{\small $^{\mathrm{c}}$ ${\rm \langle B\rangle}$ 
values are intensity-averaged mean magnitudes.}\\
{\small $^{\mathrm{d}}$ The ${\rm \langle V\rangle}$ values 
were derived scaling from the $B$ light curves according to the procedure described at the end of Section 4.2.}
\par\noindent

\end{table*}

\begin{table*}
\small
\caption[]{Number of bona fide variables plotted in the CMDs in Figs. 12, 13, and 14, divided by type.}
\label{var_type}
     $$
         \begin{array}{lccccc}
	    \hline
            \hline
           \noalign{\smallskip}
           \multicolumn{5}{c}{\rm Field~S2}\\
            \hline
            \noalign{\smallskip}
                        &{\rm RR~Lyrae~stars}&{\rm CCs}&{\rm Bin.}&{\rm Uncertain~type}\\ 
{\rm CCD~1~(upper~half)}&9                    &18	&2	    &20\\ 
{\rm [CCD~1~(lower~half)}&1                    &0	 &0	     &1{\rm ]}\\ 
{\rm CCD~2~(upper~half)}&10	      	      &14	&1	    &10\\ 
{\rm CCD~2~(lower~half)}&11	      	      &7	&0	    &21\\ 
\hline
            \noalign{\smallskip}
           \multicolumn{5}{c}{\rm Field~H1}\\
            \hline
            \noalign{\smallskip}
                        &{\rm RR~Lyrae~stars}&{\rm CCs}&{\rm Bin.}&{\rm Uncertain~type}\\ 
{\rm CCD~2~(upper~half)}&8              	&2	   &0	      &3\\ 
\hline
           \multicolumn{5}{c}{\rm Total ~numbers}\\
            \hline
            \noalign{\smallskip}
                        &{\rm RR~Lyrae~stars}&{\rm CCs}&{\rm Bin.}&{\rm Uncertain~type}\\ 
{\rm Totals}&39              	&41	   &3	      &55\\ 
\hline
\end{array}
  $$
\end{table*}


\begin{table*}
\small
      \caption[]{Distance estimates from the Classical Cepheids}
         \label{distance}
     $$
         \begin{array}{ccccccc}
	    \hline
            \hline
           \noalign{\smallskip}
         {\rm Name} &  {\rm P} &\mu PL&\mu Wes&\mu_{{\rm ogle}_{\rm HST}}&\mu PL02&\mu Wes02  \\
            \noalign{\smallskip}
            \hline
{\rm ~V2}  & 9.4   &  24.42&	24.24 &  24.52 & 24.15 &   24.42\\ 
{\rm ~V3}  & 3.25  &  24.56&    24.42 &  24.65 & 24.54 &   24.62\\
{\rm ~V4}  & 5.1   &  24.54&	24.73 &  24.63 & 24.41 &   24.92\\
{\rm ~V5}  & 2.92  &  24.37&	24.50 &  24.46 & 24.38 &   24.70\\
\hline			     
            \end{array}
	    $$
\par\noindent
\end{table*}

\clearpage

\clearpage

\begin{figure*}
\begin{center}
\includegraphics[scale=0.55, clip=true]{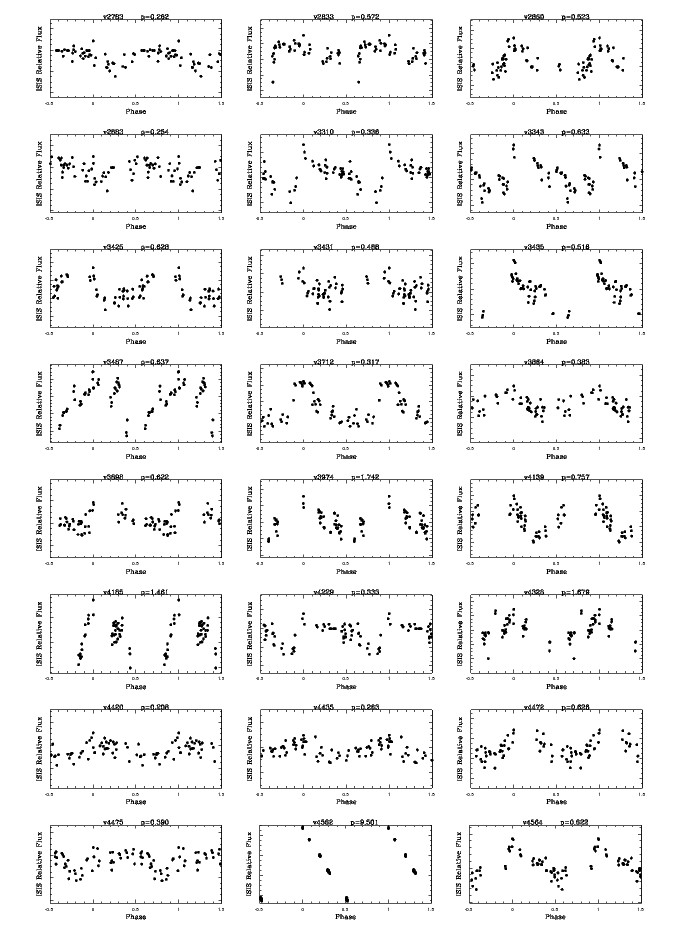}
\caption{Atlas of light curves, in $B$-band differential flux, for candidate variable stars in CCD1 of field S2. For each candidate variable
identification and a tentative period used to fold the time-series data are provided on top of the plot. Only a portion of the catalogue is 
shown here, the full atlas of light curves is published in the electronic edition of the {\it Journal}.
}
\label{light_curves_s2_chip1_p1}
\end{center}
\end{figure*}

\begin{figure*}
\begin{center}
\figurenum{9}
\includegraphics[scale=0.55, clip=true]{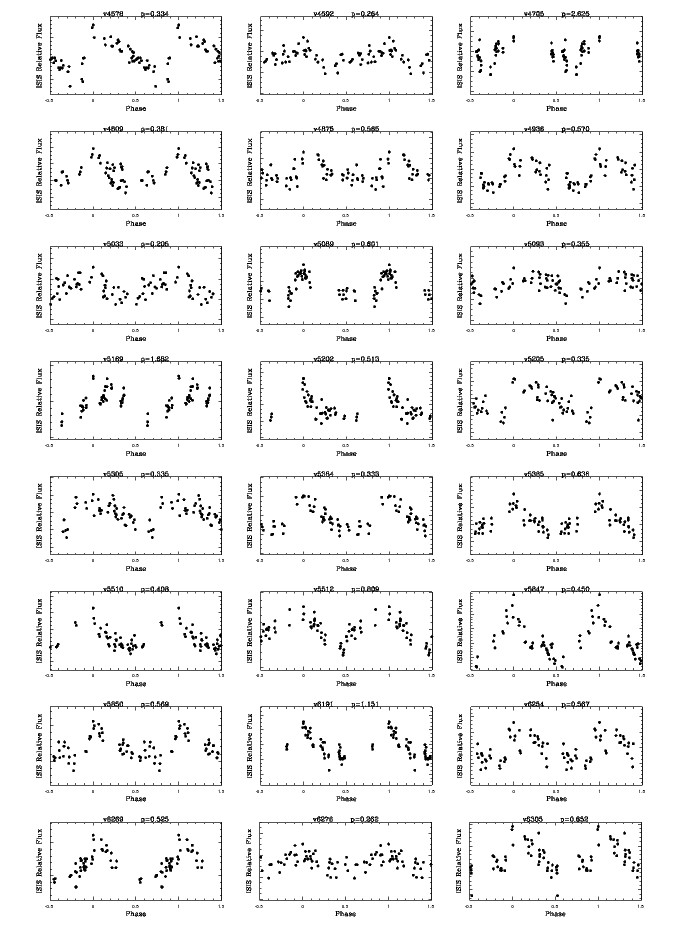}
\caption{continued
}
\end{center}
\end{figure*}
\begin{figure*}
\begin{center}
\figurenum{9}
\includegraphics[scale=0.55, clip=true]{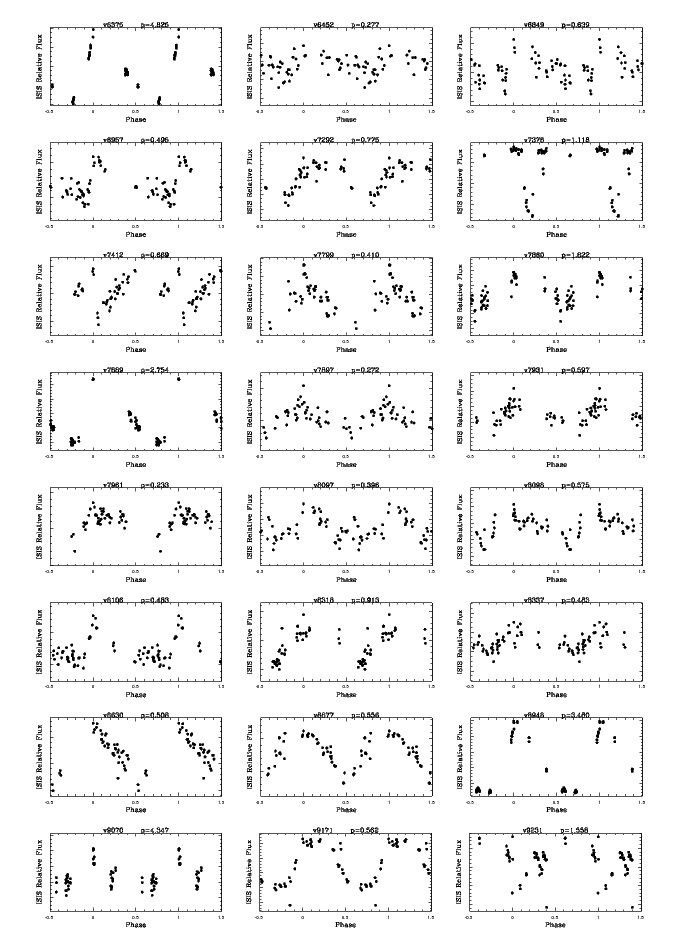}
\caption{continued
}
\end{center}
\end{figure*}
\begin{figure*}
\begin{center}
\figurenum{9}
\includegraphics[scale=0.55, clip=true]{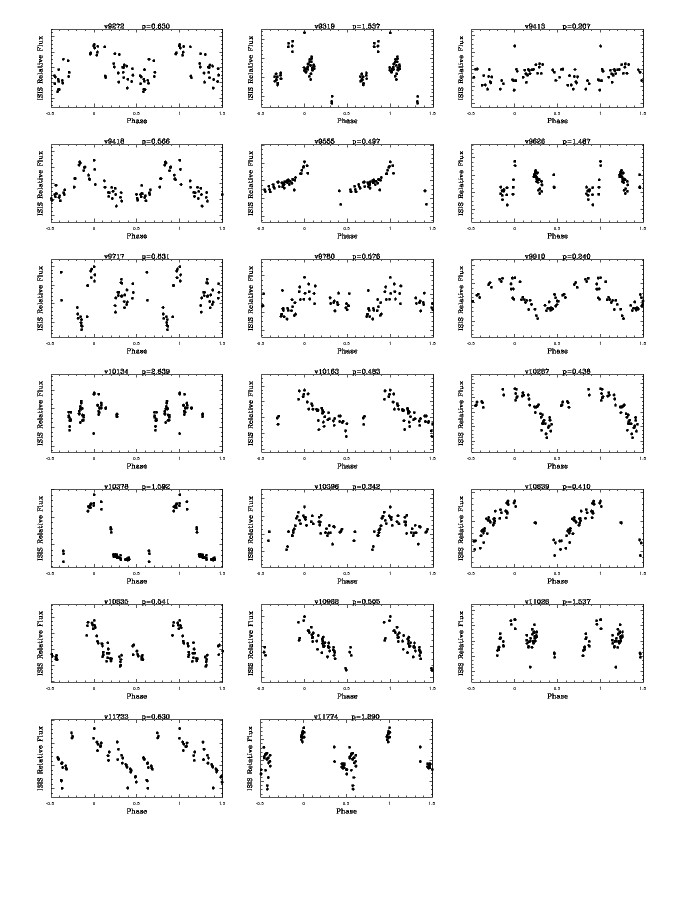}
\caption{continued
}
\end{center}
\end{figure*}

\clearpage

\begin{figure*}
\begin{center}
\includegraphics[scale=0.55, clip=true]{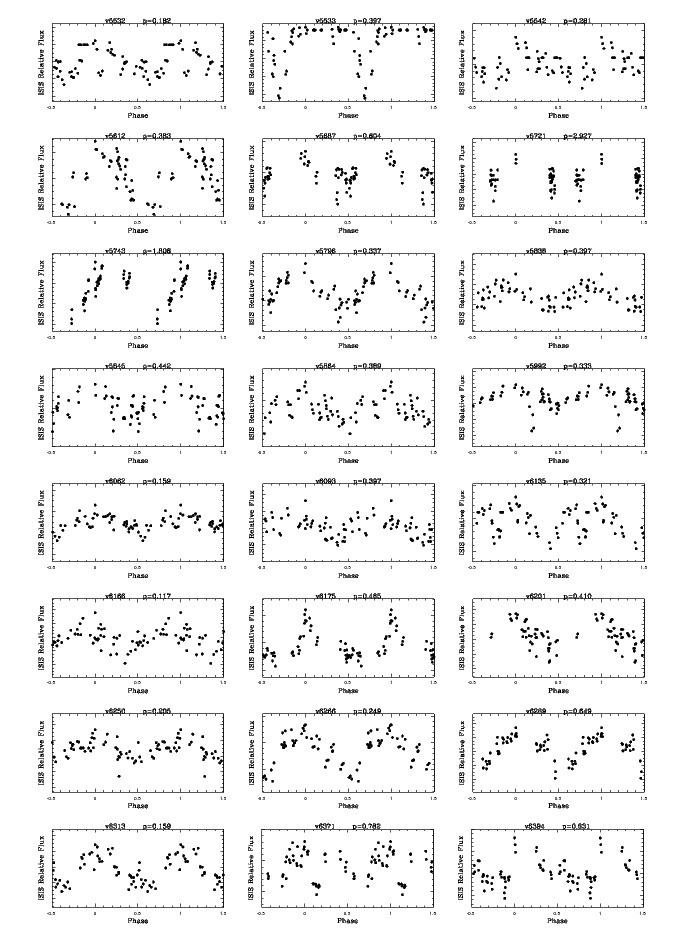}
\caption{Same as Figure~\ref{light_curves_s2_chip1_p1} for candidate variable stars in CCD2 of field S2.
}
\label{light_curves_s2_chip2_p1}
\end{center}
\end{figure*}

\begin{figure*}
\begin{center}
\figurenum{10}
\includegraphics[scale=0.55, clip=true]{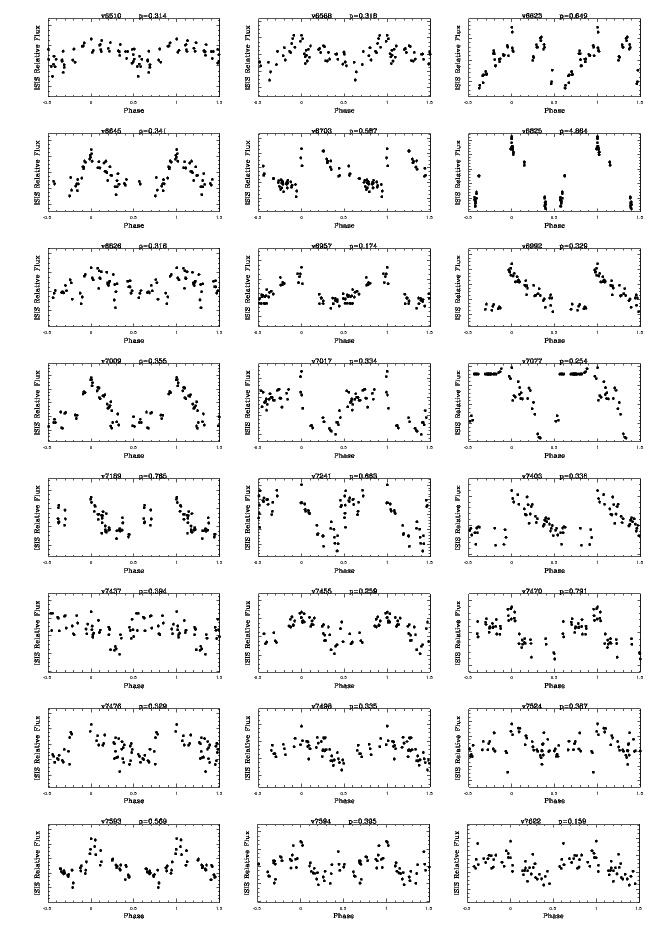}
\caption{continued
}
\end{center}
\end{figure*}

\begin{figure*}
\begin{center}
\figurenum{10}
\includegraphics[scale=0.55,  clip=true]{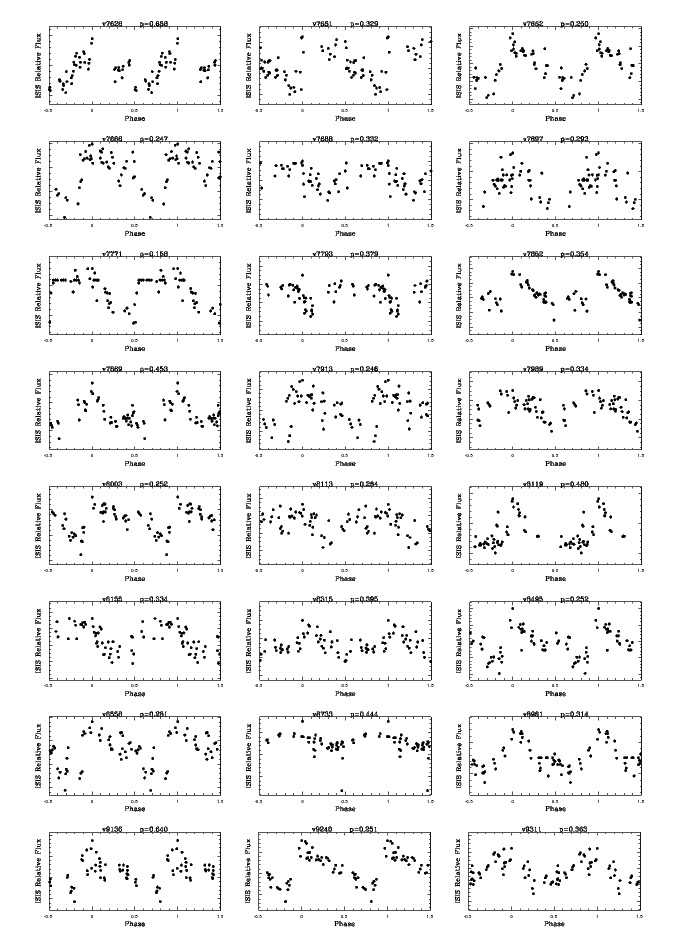}
\caption{continued
}
\end{center}
\end{figure*}

\begin{figure*}
\begin{center}
\figurenum{10}
\includegraphics[scale=0.55,  clip=true]{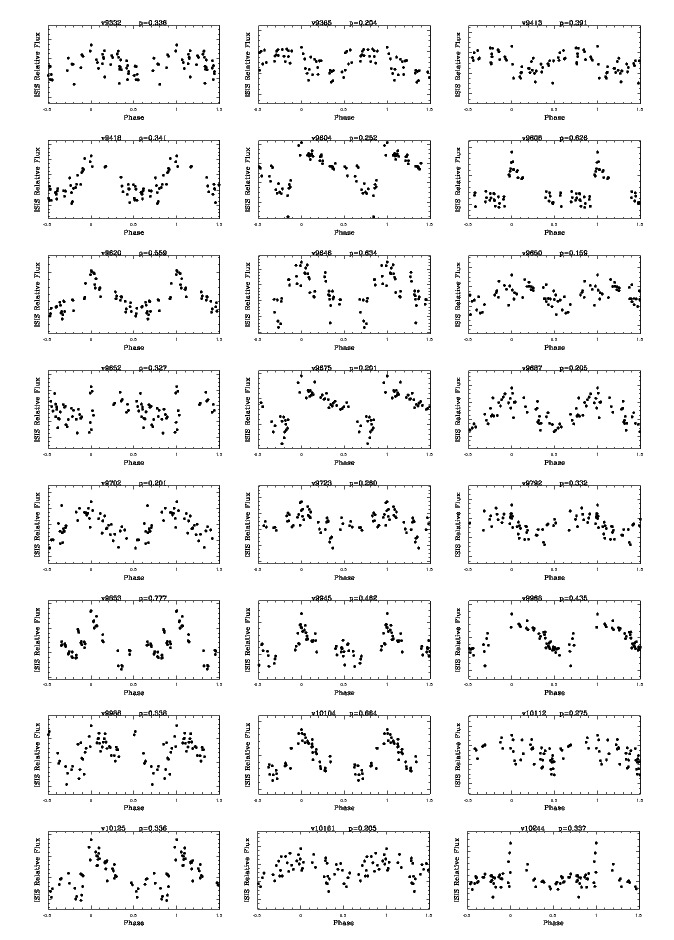}
\caption{continued
}
\end{center}
\end{figure*}

\begin{figure*}
\begin{center}
\figurenum{10}
\includegraphics[scale=0.55, clip=true]{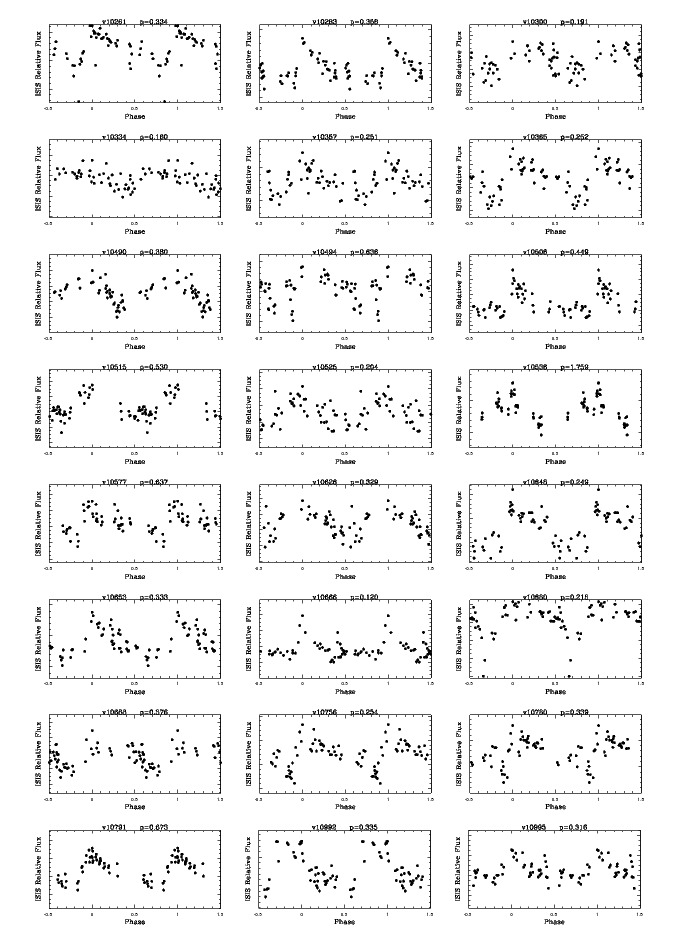}
\caption{continued
}
\end{center}
\end{figure*}

\begin{figure*}
\begin{center}
\figurenum{10}
\includegraphics[scale=0.55, clip=true]{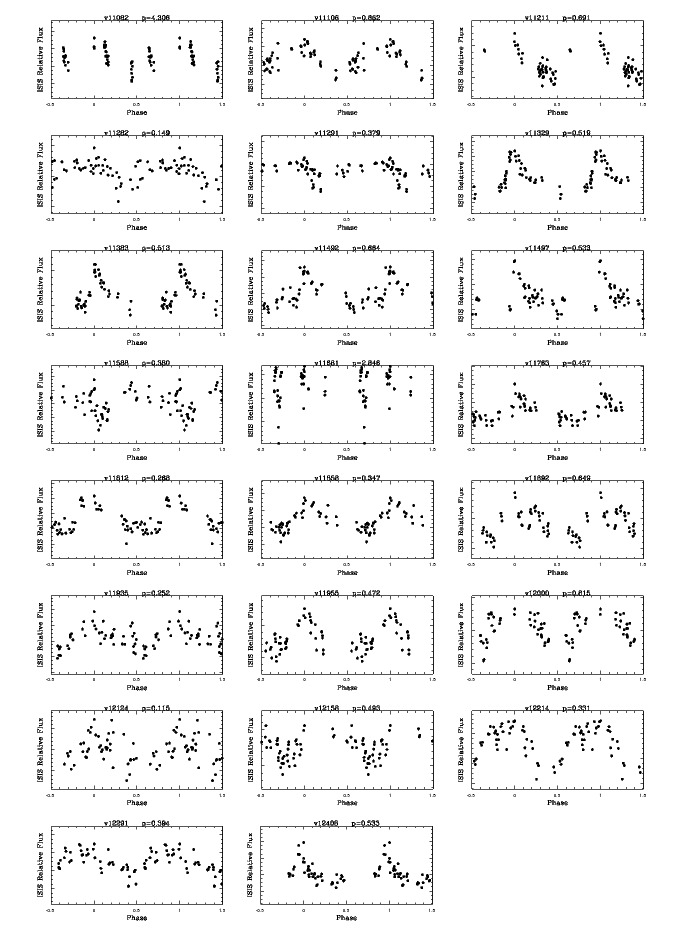}
\caption{continued
}
\end{center}
\end{figure*}

\clearpage

\begin{figure*}
\begin{center}
\includegraphics[scale=0.55, clip=true]{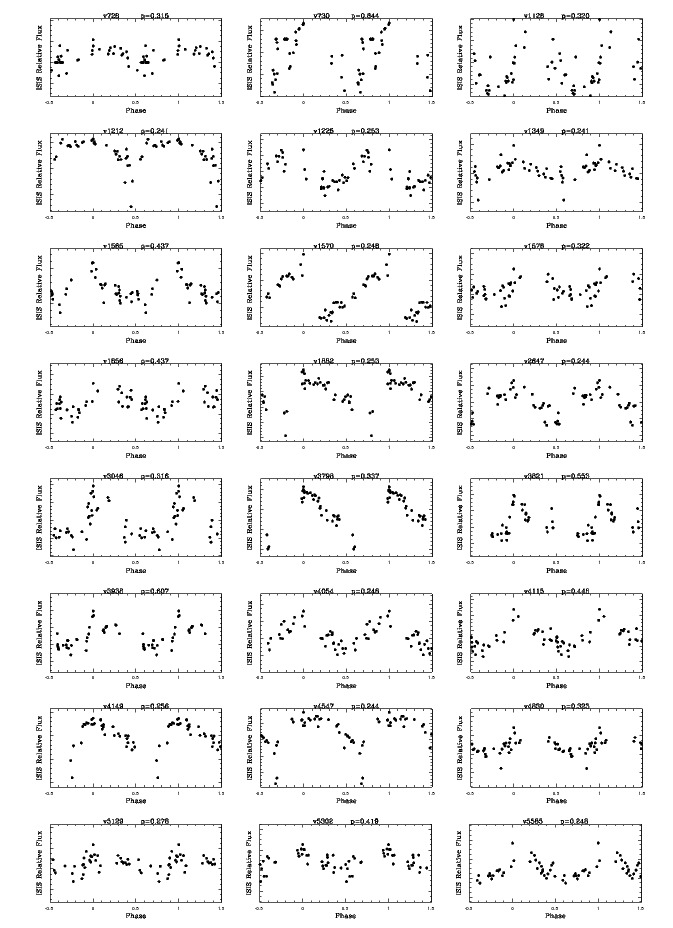}
\caption{Same as Figs.~\ref{light_curves_s2_chip1_p1} and ~\ref{light_curves_s2_chip2_p1} for candidate variable stars in CCD2 of field H1.
}
\label{light_curves_h1_chip2_p1}
\end{center}
\end{figure*}

\begin{figure*}
\begin{center}
\figurenum{11}
\includegraphics[scale=0.55,  clip=true]{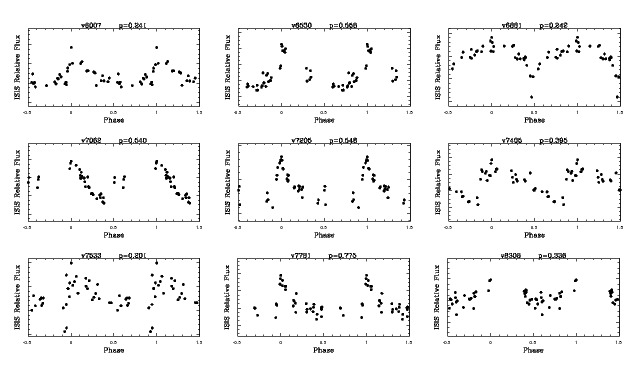}
\caption{continued
}
\end{center}
\end{figure*}

\clearpage

\begin{figure*} 
\begin{center}
\includegraphics[scale=.55]{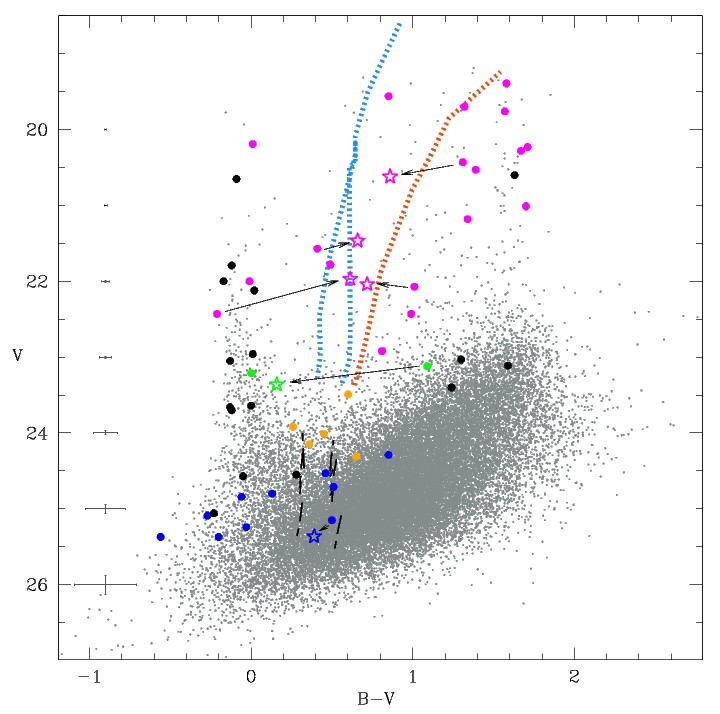}
\caption{$V,B-V$ CMD of the region of field S2 imaged on the upper portion of CCD1, with
different types of variable stars plotted by large filled dots; they include RR Lyrae stars (around $V \sim$ 25-25.4
mag, blue dots), candidate ACs/spCCs  (around $V \sim$ 24 mag, orange dots), Classical Cepheids ($V \le$ 23 mag, magenta dots), 
binary systems (green dots), and candidate variables with uncertain classification (back dots). 
All these variables are plotted according to their random phase magnitudes and colors. 
Large open stars mark 5 pulsating variables (blue and magenta stars) and 1 binary system (green star) for which we have a good sampling of
the $B$ light curves (see Fig. 8), their magnitudes and colors correspond to the average values along the full light
cycle. Arrows connect the random phase values of these 6 variables to the mean values over the full light cycle. The long-dashed black lines show the boundaries of the
theoretical instability strips for RR Lyrae stars, and for Anomalous Cepheids with
Z=0.0004 and 1.3 $<$ M$<$ 2.2 M$_{\odot}$ 
(Marconi et al. 2004). 
The dotted heavy lines represent the first overtone and fundamental blue edges (blue lines) and the fundamental 
red edge (red line) for Classical Cepheid models with Y=0.25 and 3.25 $<$ M/M$_{\odot} <$ 11 and Z=0.008 (Bono, Marconi, Stellingwerf 1999; 
Bono et al. 2002).
%
}
\label{cmd2ter}
\end{center}
\end{figure*}

\clearpage

\begin{figure*} 
\begin{center}
\includegraphics[scale=.55]{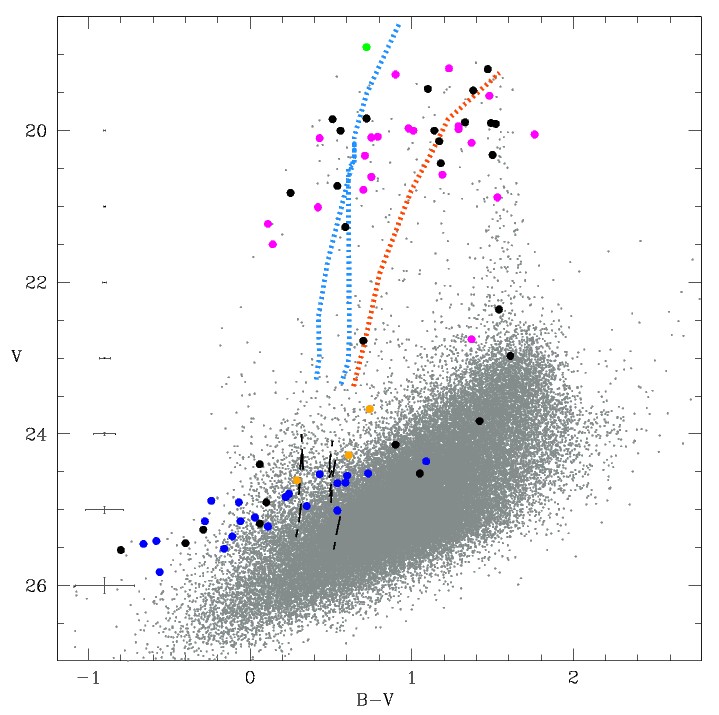}
\caption{Same as Figure~\ref{cmd2ter} for  variable stars detected in the whole CCD2 of field S2.}
%
\label{cmd2ter_s2chip2}
\end{center}
\end{figure*}

\clearpage

\clearpage

\begin{figure*} 
\begin{center}
\includegraphics[scale=.55]{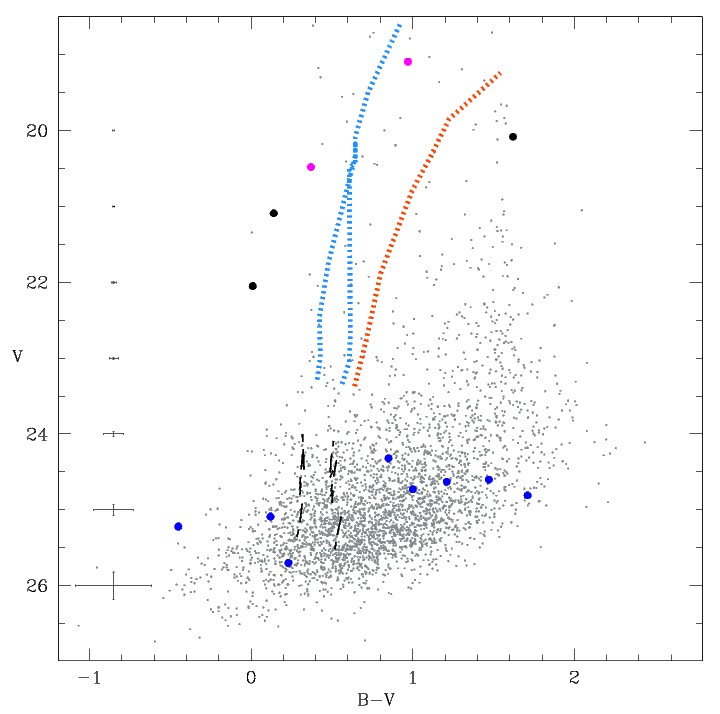}
\caption{Same as Fig.~\ref{cmd2ter} for  variable stars detected in the upper portion of CCD2 of field H1.}
%
\label{cmd2ter_h1chip2}
\end{center}
\end{figure*}

\clearpage

\begin{figure*} 
\begin{center}
\includegraphics[scale=.55]{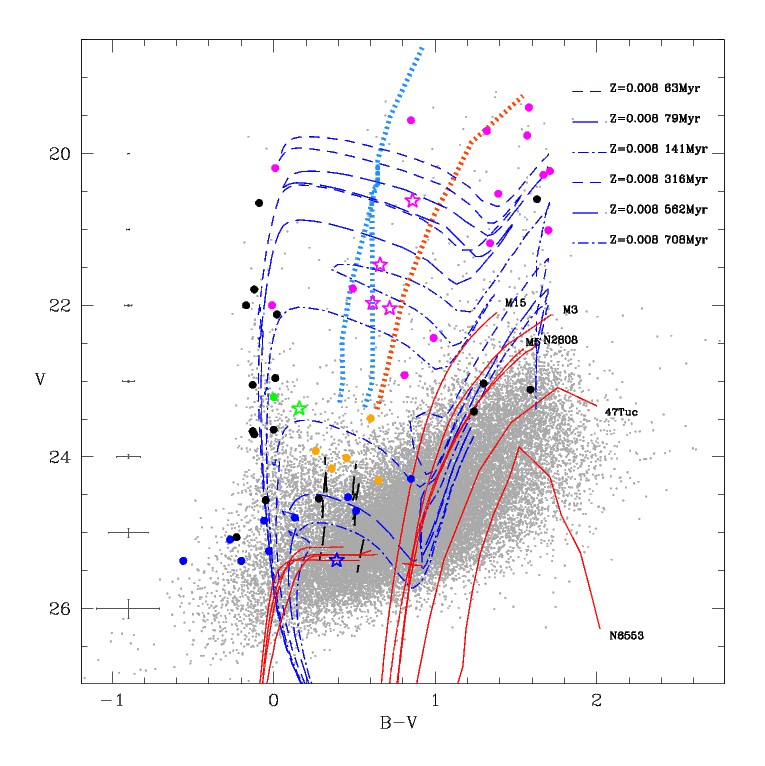}
\caption{Same as Fig.~\ref{cmd2ter} with overimposed the isochrones of Girardi et al. (2002)
for ages 63, 79, 141, 316, 562 and 708 Myr and metal content Z=0.008 (blue lines), and the mean ridge lines of the Galactic
globular clusters M15, M3, M5, NGC2808, 47 Tuc and NGC6553 (red lines).
}
\label{cmd2ter_008_gir2002_new}
\end{center}
\end{figure*}

\clearpage

\begin{figure*} 
\begin{center}
\includegraphics[scale=.55]{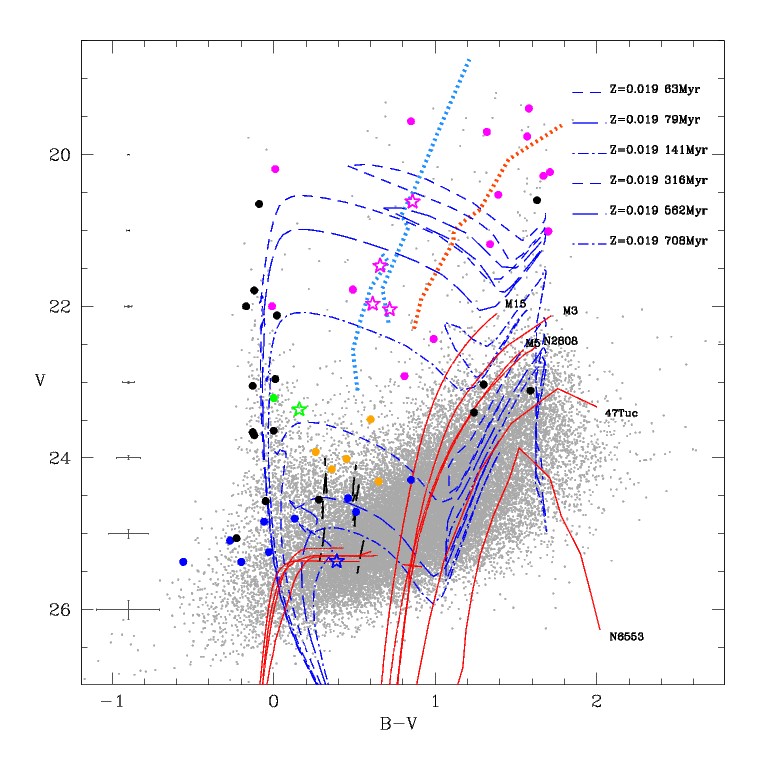}
\caption{Same as Fig.~\ref{cmd2ter_008_gir2002_new} but with isochrones and boundaries of the theoretical IS for CCs corresponding to Z=0.019.
}
\label{cmd2ter_019_gir2002_new}
\end{center}
\end{figure*}

\clearpage

\begin{figure*} 
\begin{center}
\includegraphics[scale=.55]{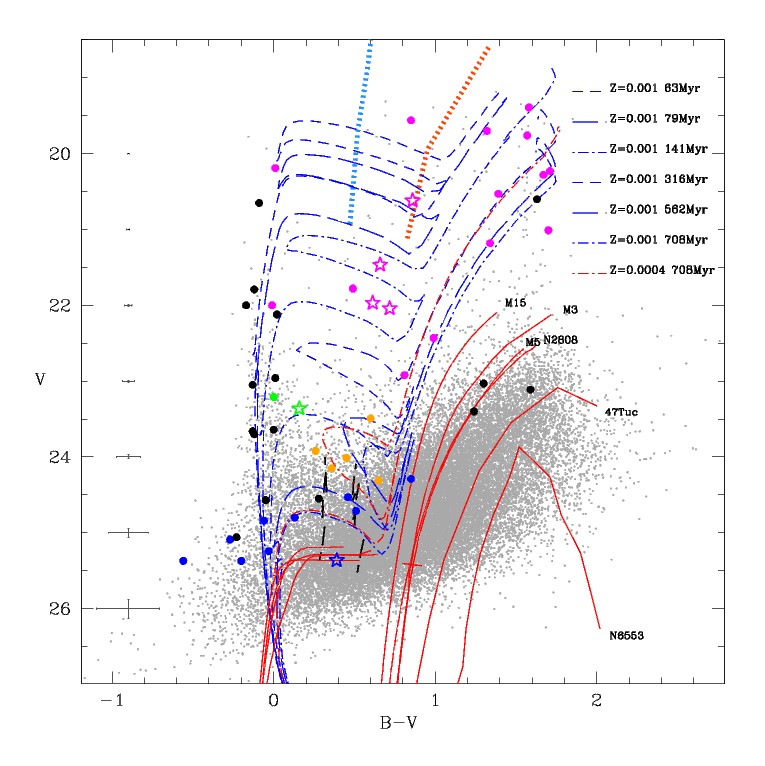}
\caption{Same as Figs.~\ref{cmd2ter_008_gir2002_new} and ~\ref{cmd2ter_019_gir2002_new} but with isochrones and boundaries of the theoretical IS 
for CCs corresponding to Z=0.001. We also show as a red dot-dashed line the isochrone with Z=0.0004 and age 708 Myr.
}
\label{cmd2ter_001_gir2002_new}
\end{center}
\end{figure*}

\end{document}